\documentclass[12pt]{article}
\usepackage{rotating}
\def\lsim{\mathrel{\rlap {\raise.5ex\hbox{$ < $}}
{\lower.5ex\hbox{$\sim$}}}}
\def\gsim{\mathrel{\rlap {\raise.5ex\hbox{$ > $}}
{\lower.5ex\hbox{$\sim$}}}}
\newcommand{\ba}{\begin{eqnarray}}
\newcommand{\ea}{\end{eqnarray}}
\newcommand{\be}{\begin{equation}}
\newcommand{\ee}{\end{equation}}
\def\ra{\rightarrow}
\def\lra{\leftrightarrow}
\def\ze{\zeta}

\begin{document}
\begin{titlepage}
\begin{flushright}

hep-th/9901098
 {\hskip.5cm}\\
\end{flushright}
\begin{centering}
\vspace{.3in}
{\bf $N=1$  supersymmetric $SU(4)\times SU(2)_L\times SU(2)_R$
     effective theory    \\
from the weakly coupled heterotic superstring}\\
 \vspace{2 cm}
{G.K. Leontaris and J. Rizos} \\ \vskip 1cm

{\it {Physics Department, University of Ioannina\\
Ioannina, GR45110, GREECE}}\\

\vspace{1.5cm}
{\bf Abstract}\\
\end{centering}
\vspace{.1in}
In the context of the free--fermionic formulation of
the heterotic superstring, we construct a three generation $N=1$
supersymmetric $SU(4)\times SU(2)_L\times SU(2)_R$ model
supplemented by an $SU(8)$ hidden gauge symmetry and five Abelian
factors. The symmetry breaking to the standard model is achieved
using vacuum expectation values of a Higgs pair in
$({\bf4},{\bf2}_R)+({\bf\bar 4},{\bf2}_R)$ at a high scale. One
linear combination of the Abelian symmetries is anomalous and is
broken by vacuum expectation values of singlet fields along the
flat directions of the superpotential. All consistent string vacua
of the model are completely classified by solving the
corresponding system of $F-$ and $D-$flatness equations including
non--renormalizable terms up to sixth order. 
The requirement of existence of electroweak massless doublets 
further restricts the phenomenologically viable vacua.
The third generation fermions receive masses from the tree--level
superpotential. Further, a complete calculation of all
non--renormalizable fermion mass terms up to fifth order shows
that in certain string vacua the hierarchy of the fermion families
is naturally obtained in the model as the second and third
generation fermions earn their mass from fourth and fifth order
terms. Along certain flat directions it is shown that the ratio of
the $SU(4)$ breaking scale and the reduced Planck mass is equal to
the up quark ratio $m_c/m_t$ at the string scale. An additional
prediction of the model, is the existence of a  $U(1)$ symmetry
carried by the fields of the hidden sector, ensuring thus the
stability of the lightest hidden state. It is proposed that the
hidden states may account for the invisible matter of the
universe.
 \vspace{2cm}
\begin{flushleft}
January 1999
\end{flushleft}
\hrule width 6.7cm \vskip.1mm{\small \small}
 \end{titlepage}
\section{Introduction}

During the last decade, a lot of work has been devoted in the
construction of effective low energy models of elementary
particles from the heterotic superstring. Several old successful
$N=1$ supersymmetric grand unified theories (GUTs) have been
recovered through the string approach
\cite{sm,gkmr,aehn,AL,ALR,more}, however only few of them were
able to rederive a number of successful predictions of their
predecessors. Yet, new avenues and radical ideas that were
previously not considered or only poorly explored, have now been
painstakingly investigated in the context of string derived or
even string inspired effective theories. Among them, the issue of
the additional $U(1)$--symmetries which naturally appear in string
models and the systematic derivation of non--renormalizable terms
boosted our understanding of the observed mass hierarchies and
impelled people to systematically classify all possible textures
consistent with the low energy phenomenology. Further, new and
astonishingly simpler mechanisms of GUT symmetry breaking were
introduced due to the absence of large Higgs representations, at
least in the simplest Kac--Moody level ($k=1$) string
constructions.

In addition to the above good omen, some embarrassing difficulties
have also appeared, such as the existence of unconfined
fractionally charged states -- which belong to representations not
incorporated in the usual GUTs -- and the very high unification
scale. The new representations come as a result of the breaking of
the large string symmetry via the GSO projections. The appearance
of such states are not necessarily an ominous warning for a
particular model, although a mechanism should be invented to make
them disappear from the light spectrum. The real major difficulty
however, was the generic property of the high string scale in
contrast to the usual supersymmetric GUTs which unify at about two
orders of magnitude below the string mass. In the weakly coupled
heterotic string theory, this problem can find a solution in
specific models, when extra matter multiplets exist to properly
modify the running of the gauge couplings, or possible
intermediate symmetries and string threshold effects \cite{thres}
can help gauge couplings converge to their experimentally
determined values at low energies.

In this paper, we derive an improved version of a string model
proposed in~\cite{ALR}, based on the observable gauge symmetry
$SO(6)\times O(4)$ (isomorphic to $SU(4)\times SU(2)_L \times
SU(2)_R$ Pati--Salam (PS) gauge group \cite{PS}) in the context of
the free--fermionic formulation of the four dimensional
superstring. As shown in \cite{AL} this gauge symmetry breaks down
to the standard model without the use of the adjoint or any higher
representation thus it can be built directly at the $k=1$
Kac--Moody level. (Higher Kac--Moody level models are also
possible to build, however, they imply small unification scale
values of $\sin^2\theta_W$ \cite{schell}.)

The models based on the PS gauge symmetry have also certain
phenomenological advantages. Among them, is the absence of
coloured gauge fields mediating proton decay. This fact allows for
the possibility of having a low $SU(4)$ breaking scale compared to
that of other GUTs, provided that the Higgs coloured fields do not
have dangerous Yukawa couplings with ordinary matter. Possible
ways to avoid fast proton decay have been discussed also recently
in the literature \cite{bwgrt}.

Moreover, specific GUT relations among the Yukawa couplings, like
the bottom--tau equality, give successful predictions at low
energies, while at the same time such relations reduce the number
of arbitrary Yukawa couplings even in the field--theory version.

The above nice features are exhibited in the present string
construction. Using a variant of the original string basis and the
GSO projection coefficients \cite{ALR}, we obtain an effective
field theory model with three generations and exactly one Higgs
pair to break the $SU(4)\times SU(2)_R$ gauge symmetry. The
effective low energy theory is the $N=1$ supersymmetric
$SU(3)\times SU(2)\times U(1)$ electroweak standard gauge symmetry
broken to $SU(3)\times U(1)_{em}$ by the two Higgs doublet fields.
We derive the complete massless spectrum of the model and the
Yukawa interactions including non--renormalizable terms up to
sixth order. Among the massless states, a mirror (half)--family is
also obtained which acquires mass at a very large scale. There are
also singlet fields and exotic doublet representations with a
sufficient number of Yukawa couplings. All the observable and
hidden fields appear with charges under five surplus $U(1)$
factors where one linear combination of them is anomalous. The
anomalous $U(1)$ symmetry generates a $D$--term contribution,
which can be cancelled if some of the singlet fields acquire
non--zero vacuum expectation values (vevs). To find the true
vacua, we solve the $F-$ and $D-$flatness conditions and classify
all possible solutions involving observable fields with non--zero
vevs. We analyze in detail three characteristic cases where
superpotential contributions up to sixth order suffice to provide
fermion mass terms for all generations. Phenomenologically
interesting alternative solutions are also proposed in the case
where some of the hidden fields develop vevs too.

Among the novel features of the present model is the existence of
a $U(1)$ symmetry -- carried by hidden and exotic fields -- which
remains unbroken. As a consequence, the lightest hidden state is
stable. These states form various potential mass terms in the
superpotential of the model. In our analysis, we show the
existence of proper flat directions where the lightest state
obtains a mass at an intermediate scale, leading to interesting
cosmological implications.

The paper is organized as follows: In Section 2 we give a brief
description of the supersymmetric version of the model and discuss
various phenomenological features, including the economical Higgs
mechanism, the mass spectrum and the renormalization group. In
particular we show how the Pati--Salam symmetry dispenses with the
use of Higgs fields in the adjoint of $SU(4)$ to break down to the
standard model. In Section 3, we propose the string basis as well
as the GSO projections which yield the desired gauge symmetry and
the massless spectrum. In Section 4 the gauge symmetry breaking of
the string version is analyzed. Moreover, due to the existence of
additional $U(1)$ symmetries, the issue of new (non--standard)
hypercharge embeddings is discussed. Particular embeddings where
all fractionally charged states obtain integral charges are
discussed in some detail. In Section 5 we derive the
superpotential couplings and present a preliminary
phenomenological analysis to set the low energy constraints and
reduce the number of phenomenologically acceptable string vacua.
In Section 6 we classify all solutions of the $F$-- and
$D$--flatness equations including non--renormalizable
superpotential contributions up to sixth order. A detailed
phenomenological analysis of the promising string vacua in
connection with their low energy predictions is presented in
Section 7. Particular attention is given in the doublet higgs mass
matrix, the fermion mass hierarchy and the colour triplet mass
matrix. In Section 8, we present a brief discussion on the role of
the hidden sector and extend the solutions of string vacua
including hidden field vevs. Finally, in the Appendices A--D we
present tables with the complete string spectrum, details about
the derivation of the higher order non--renormalizable
superpotential terms, the $D$-- and $F$-- flatness equations with
non-renormalizable contributions and the complete list of their
tree--level solutions.


\section{The Supersymmetric $SU(4)\times SO(4)$  Model \label{model}}

There is a minimal supersymmetric $SU(4)\times O(4)$ Model which
can be considered as a surrogate effective GUT of the possible
viable string versions, incorporating all the basic features of a
phenomenologically viable string model. The Yukawa couplings are
determined by the Pati--Salam (PS) gauge symmetry and possible
additional $U(1)$--family symmetries which are usually added (as
in any other GUT) by phenomenological requirements, (i.e., fermion
mass hierarchy, proton stability etc). This GUT version, however,
provides us with insight in constructing the fully realistic
string version. Therefore, here we briefly summarize the parts of
the model relevant for our analysis~\cite{AL}. The gauge group is
$SU(4)\times O(4)$, or equivalently the  PS gauge symmetry
\cite{PS}
\begin{equation}
\mbox{SU(4)}\times \mbox{SU(2)}_L \times \mbox{SU(2)}_R.
\label{422}
\end{equation}
The left--handed quarks and leptons are accommodated in the
following representations,
\begin{equation}
{F^i}^{\alpha a}_L=({\bf4},{\bf2},{\bf1})= \left(\begin{array}{cc}
 u^\alpha & \nu \\ d^\alpha & e
\end{array} \right)^i
\end{equation}
\begin{equation}
{\bar{F}}_{x \alpha R}^i=({\bf\bar{4}},{\bf1},{\bf{2}})=
\left(\begin{array}{cc} {d}^{c\alpha}  & e^c  \\ {u}^{c\alpha} &
{\nu^c}
\end{array} \right)^i
\end{equation}
where $\alpha =1,\ldots ,4$ is an $SU(4)$ index, $a,x=1,2$ are
SU(2)$_{L,R}$ indices, and $i=1,2,3$ is a family index.  The Higgs
fields are contained in the following representations,
\begin{equation}
h_{a}^x=({\bf1},{\bf2},{\bf2})= \left(\begin{array}{cc}
  {h_+}^u & {h_0}^{d} \\ {h_0}^u & {h_-}^d \\
\end{array} \right) \label{h}
\end{equation}
where $h^d$ and $h^u$ are the low energy Higgs superfields
associated with the minimal supersymmetric standard model (MSSM).
The two `GUT' breaking higgs representations are
\begin{equation}
{H}^{\alpha b}=({\bf4},{\bf1},{\bf2})= \left(\begin{array}{cc}
u_H^\alpha & \nu_H
\\ d_H^\alpha& e_H
\end{array} \right) \label{H}
\end{equation}
and
\begin{equation}
{\bar{H}}_{\alpha x}=({\bf\bar{4}},{\bf1},{\bf{2}})=
\left(\begin{array}{cc} u_H^{c\alpha} & e_H^c \\ {u}_H^{c\alpha} &
{\nu}_H^c
\end{array} \right). \label{barH}
\end{equation}
{}Fermion generation multiplets transform to each other under the
changes ${\bf4}\ra {\bf\bar 4}$ and ${\bf2}_L\ra {\bf2}_R$ while
the bidoublet higgs multiplet transforms to itself. However, the
pair of fourplet--higgs fields does not have this property,
discriminating  ${\bf2}_L$ and ${\bf2}_R$. Thus, when they develop
vevs along their neutral components $\tilde{\nu}_H,
\tilde{{\nu}}_H^c$,
\begin{equation}
\langle H\rangle=\langle\tilde{\nu}_H\rangle\sim M_{GUT}, \ \
\langle\bar{H}\rangle=\langle\tilde{{\nu}}_H^c\rangle\sim M_{GUT}
\label{HVEV}
\end{equation}
they break the $SU(4)\times SU(2)_R$ part of the gauge group,
leading  to the standard model symmetry at $M_{GUT}$
\begin{equation}
\mbox{SU(4)}\times \mbox{SU(2)}_L \times \mbox{SU(2)}_R
\longrightarrow
\mbox{SU(3)}_C \times \mbox{SU(2)}_L \times \mbox{U(1)}_Y.
\label{422to321}
\end{equation}
Under the symmetry breaking in Eq. (\ref{422to321}), the bidoublet
Higgs field $h$ in Eq. (\ref{h}) splits into two Higgs doublets
$h^u$, $h^d$ whose neutral components subsequently develop weak
scale vevs,
\begin{equation}
\langle h_0^d\rangle =v_1, \ \ \langle h_0^u\rangle =v_2 \label{vevs1}
\end{equation}
with $\tan \beta \equiv v_2/v_1$.

In addition to the Higgs fields in Eqs.~(\ref{H}),(\ref{barH}) the
model also involves an $SU(4)$ sextet field
$D=({\bf6},{\bf1},{\bf1})$ and four singlets $\phi_0$ and
$\varphi_i$, $i= 1,2,3$. $\phi_0$ is going to acquire a vev of the
order of the electroweak scale in order to realize the Higgs
doublet mixing, while $\varphi_i$ will participate in an extended
`see--saw' mechanism to obtain light majorana masses for the
left--handed neutrinos. Under the symmetry property
$\varphi_{1,2,3}\ra (-1)\times \varphi_{1,2,3}$ and $H
(\bar{H})\ra (-1)\times H (\bar{H})$ the tree--level mass terms of
the superpotential of the model read \cite{AL}:
\begin{equation}
W =\lambda^{ij}_1F_{iL}\bar{F}_{jR} h
+\lambda_2HHD+\lambda_3\bar{H}\bar{H}D+\lambda^{ij}_4H\bar{F}_{jR}
\varphi_i +\mu\varphi_i\varphi_j+\mu hh \label{W}
\end{equation}
where $\mu = \langle\phi_0\rangle\sim {\cal O}(m_W)$.
The last term generates the higgs mixing between the two
SM Higgs doublets in order to prevent the appearance of a massless
electroweak axion. The following decompositions take place under the
symmetry breaking (\ref{422to321}):
\ba
    {F}_L({\bf4},{\bf2},{\bf1})     &\ra &
     Q({\bf3},{\bf2},-\frac 16) + \ell{({\bf1},{\bf2},\frac 12)}
     \nonumber\\
\bar{F}_R({\bf\bar 4},{\bf1},{\bf2})&\ra &  u^c({\bf\bar 3},
{\bf1},\frac 23)+d^c({\bf\bar 3},{\bf1},-\frac 13)+
                           \nu^c({\bf1},{\bf1},0)+ e^c({\bf1},
                           {\bf1},-1)
                           \nonumber\\
\bar{H}({\bf\bar 4},{\bf1},{\bf2})&\ra &  u^c_H({\bf\bar 3},
{\bf1},\frac 23)+d^c_H({\bf\bar 3},{\bf1},-\frac 13)+
                            \nu^c_H({\bf1},{\bf1},0)+
                             e^c_H({\bf1},{\bf1},-1)
                             \nonumber\\
{H}({\bf4},{\bf1},{\bf2})&\ra &  u_H({\bf3},{\bf1},-\frac
23)+d_H({\bf3},{\bf1},\frac 13)+
              \nu_H({\bf1},{\bf1},0)+ e_H({\bf1},{\bf1},1)
              \nonumber\\
D({\bf6},{\bf1},1)  &\ra &  D_3({\bf3},{\bf1},\frac 13) +
\bar{D}_3({\bf\bar 3},{\bf1},-\frac 13)
                             \nonumber\\
h({\bf1},{\bf2},{\bf2})
 &\ra &  h^d({\bf1},{\bf2},\frac 12) + h^u({\bf1},{\bf2},-\frac 12)
 \nonumber
\ea
where the fields on the left appear with their quantum numbers under the
PS gauge symmetry, while the fields on the right are shown with their
quantum numbers under the SM symmetry.

The superpotential Eq. (\ref{W}) leads to the following neutrino
mass matrix \cite{AL}
\begin{eqnarray}
{\cal M}_{\nu,\nu^c,\varphi} = \left(\begin{array}{ccc}
0 & m^{ij}_u & 0 \\
m^{ji}_u& 0 & M_{GUT} \\
0  & M_{GUT}  & \mu \\ \end{array}\right)
\label{se-sa}
\end{eqnarray}
in the basis $(\nu_i , {\nu}_j^c ,\varphi_k)$. Diagonalization of
the above gives three light neutrinos with masses of the order
$\mu(m_u^{ij}/M_{GUT})^2$ as required by the low energy data, and
leaves right--handed majorana neutrinos with masses of the order
$M_{GUT}$.  Additional terms not included in Eq. (\ref{W}) may be
forbidden by imposing suitable discrete or continuous symmetries
\cite{Qaisar,BPW} which, in fact, mimic the role of various $U(1)$
factors and string selection rules appearing in realistic string
models. The sextet field $D({\bf 6},{\bf1},{\bf1})$ carries colour,
 while after
the symmetry breaking it decomposes in a triplet/triplet--bar pair
with the same quantum numbers of the down quarks. Now, the terms
in Eq. (\ref{W}) $HHD$ and $\bar{H} \bar{H}D$ combine the uneaten
(down quark--type) colour triplet parts of $H$, $\bar{H}$ with
those of the sextet $D$ into acceptable GUT--scale mass terms
\cite{AL}. When the $H$ fields attain their vevs at
$M_{GUT}\sim10^{16}$ GeV, the superpotential of Eq. (\ref{W})
reduces to that of the MSSM augmented by right-handed neutrinos.
Below $M_{GUT}$ the part of the superpotential involving matter
superfields is just
\begin{equation}
W
=\lambda^{ij}_UQ_i{u^c}_jh_2+\lambda^{ij}_DQ_i{d^c}_jh_1 n
 +\lambda^{ij}_E\ell_i{e^c}_jh_1+ \lambda^{ij}_NL_i{\nu}_j^ch_2 +
\cdots
\label{MSSMmatter}
\end{equation}
The Yukawa couplings in Eq. (\ref{MSSMmatter}) satisfy the
boundary conditions
\begin{equation}
\lambda^{ij}_1 (M_{GUT}) \equiv \lambda^{ij}_U(M_{GUT})
= \lambda^{ij}_D (M_{GUT})=
\lambda^{ij}_E(M_{GUT}) = \lambda^{ij}_N(M_{GUT}).
 \label{boundary}
\end{equation}
Thus, Eq. (\ref{boundary}) retains the successful relation
$m_{\tau}=m_b$ at $M_{GUT}$. Moreover from the relation
$\lambda^{ij}_U(M_{GUT}) = \lambda^{ij}_N(M_{GUT})$, and the
fourth term in Eq. (\ref{W}), through the see--saw mechanism we
obtain light neutrino masses which satisfy the experimental
limits. The $U(1)$ symmetries imposed by hand in this simple
construction play the role of family symmetries $U(1)_A$, broken
at a scale $M_A>M_{GUT}$ by the vevs of two $SU(4)\times O(4)$
singlets $\theta,\bar\theta$, carrying charge under the
family symmetries and leading to operators of the form
\begin{equation} \label{newop}
O_{ij}\sim (F_i\bar{F}_j)h\left(\frac{H\bar{H}}{M^2}\right)^r
\left(\frac{\theta^n \bar{\theta}^m}{{M'}^{n+m}}\right)+
{\mbox h.c.}
\label{newoperators}
\end{equation}
obtained from  non--renormalizable (NR) contributions
to the superpotential. Here, $M'$
represents a high scale $M' > M_{GUT}$ which may be identified either
with the $U(1)_A$ breaking scale $M_A$ or with the string scale
$M_{string}$. Such terms have the task of filling in the entries of
fermion mass matrices, creating textures with a hierarchical mass
spectrum and mixing effects between the fermion generations.

Before we proceed to the construction of a particular string model
let us examine how a three--generation $SU(4)\times SU(2)_L\times
SU(2)_R$ model can be realized. As we have already explained the
fermion generations are accommodated in
$F_L({\bf4},{\bf2},{\bf1})+{\bar F}_R({\bf\bar4},{\bf1},{\bf2})$
while the Higgs fields are accommodated in
$F_R({\bf4},{\bf1},{\bf2})+{\bar F}_R ({\bf\bar4},{\bf1},{\bf2})$
representations. In the free--fermionic formulation the
$SU(2)_L\times SU(2)_R$ is realized as $O(4)$ and the ${\bf2}_L$ and
${\bf2}_R$ representations are the two spinor representations
(${\bf2}^{\pm}$) of $O(4)$. Calling $n_{+}, n_{-}, {\bar n_{+}}, {\bar
n_{-}}$ the number of
$({\bf4},{\bf2}^{+}),({\bf4},{\bf2^{-}})$,$({\bf\bar4},{\bf2^{+}})$
and $({\bf{\bar4}},{\bf2^{-}})$ representations we come to the
conclusion that one minimal three generation model is obtained for
$${\vec n}_{min}=\left(n_{+}, n_{-}, {\bar n_{+}}, {\bar
n_{-}}\right) =(3,1,0,4)$$ where ${\bf2}_L$ is identified with ${\bf2}^{+}$.
A mirror minimal model can be obtained by interchanging the two
$SU(2)$'s, i.e. $L\leftrightarrow R$, and identifying ${\bf2}^{+}\sim
{\bf2}_R$
$${\vec n}_{min}=\left(n_{+}, n_{-}, {\bar n_{+}}, {\bar
n_{-}}\right) =(1,3,4,0).$$
 {}Furthermore, one can consider the existence of vector--like
 states that do not affect the net number of generations since,
 in principle they can obtain superheavy
masses. Thus a general three--generation $SU(4)\times
SU(2)_L\times SU(2)_R$ model corresponds to one the following
vectors $${\vec n}_{rs}=\left(n_{+}, n_{-}, {\bar n_{+}}, {\bar
n_{-}}\right) =(3+r,1+s,r,4+s)\ , \ r,s=0,1,\dots$$ or $${\vec
n}_{rs}=\left(n_{+}, n_{-}, {\bar n_{+}}, {\bar n_{-}}\right)
=(1+s,3+r,4+s,r)\ , \ r,s=0,1,\dots$$ We can rewrite the above
relations in a more compact form
\begin{eqnarray}
n_{+}+n_{-}&=&{\bar n}_{+}+{\bar n}_{-}=4+p, \
p=0,1,2,\dots\nonumber\\ n_{+}-{\bar n}_{+}&=&{\bar
n}_{-}-n_{-}=\pm3 \label{lr}
\end{eqnarray}
Thus, there exist an infinity of three--generation $SU(4)\times
O(4)$ $\sim$ $SU(4)\times SU(2)_L\times SU(2)_R$ models each of
them uniquely characterized by an integer ($p$) related to the
differences (\ref{lr}) and a sign ($\pm$). We will therefore refer
to a particular model using the notation $k^{\pm}$ that is the two
minimal models will be referred as $0^{+}$ and $0^{-}$.

As stressed in the introduction, one severe problem that has to be
resolved in a candidate string model is the discrepancy between
the unification scale as this is found when the minimal
supersymmetric spectrum is considered, and the two orders higher
string scale implied by theoretical calculations. In previous
works, it was shown that this difficulty may be overcome in
several ways \cite{alt,sk}. In particular, the class of string
models as that of Ref. \cite{ALR} predict additional matter fields
which can help the couplings merge at the high string scale
without disturbing the low energy values of $\sin^2\theta_W$ and
$\alpha_s$. Perhaps the most elegant way to achieve this, is to
make the couplings run closely from the string to the
phenomenological unification scale $M_U\sim 10^{16}$GeV. As a
first step one may add the mirror fields \cite{sk} \ba M =
({\bf4},{\bf2},{\bf1}); & \bar{M} = ({\bf\bar 4},{\bf2},{\bf1})
\label{mir}
\ea which guarantee the equality of the $SU(2)_L$ and $SU(2)_R$
gauge couplings $g_L=g_R$ between the two scales. According to the
classification proposed to the previous paragraph, this model is
classified as $1^+$ or $1^-$. The running of the $SU(4)$ coupling
can be adjusted by an additional number of extra colour sextets
which are in general available in the string versions of the
present model. Indeed, with three generations and denoting
collectively the number of fourplet sets with $n_4$ the beta
functions now become \ba b_{2L}\equiv b_{2R} = -1- 4 n_4; & b_4 =
6-n_6 -4 n_4 \ea which show that a sufficient number of sextet
fields may guarantee a $g_4$ running almost identical with that of
$g_{L,R}$. The string model we are proposing in the next section
has exactly one mirror pair and four sextet fields, whereas
additional exotic states may also contribute to the beta functions
if they remain in the light spectrum.

The introduction of the mirror representations (\ref{mir}) leads
to the existence of another symmetry in the model: we observe that
the whole spectrum  now is completely symmetric with respect to
the two $SU(2)$'s in the sense that  under the simultaneous change
${\bf2}_L\lra {\bf2}_R$  and the ${\bf4}\lra {\bf\bar{4}}$ of $SU(4)$, 
it remains
invariant. More precisely, under this symmetry the representations
of the model are mapped as follows \ba \bar{F}_{R}& \lra &F_L
\nonumber\\ H,\bar{H} &\lra& \bar{M}, M\\ D, h,\phi_i &\lra &D, h,
\phi_i\nonumber \label{msym} \ea This symmetry persists also in
the present string model, while tree--level as well as higher
order Yukawa interactions are also invariant under these changes.
As we will see, this symmetry is broken by the vacuum which will
be determined by the specific solutions of the flatness conditions.

After the above short description, we are ready to present
the string derived model where most of the above features
appear naturally. In addition, novel predictions will emerge
 such as the appearance of exotic states with
charges which are fractions of those of ordinary quarks and leptons,
a hidden `world' and a low energy $U(1)$ symmetry.

\section{The String Model}

In the four dimensional free--fermionic formulation of the
heterotic superstring, fermionic degrees of freedom on the world
sheet are introduced to cancel the conformal anomaly. The
right--moving non--supersymmetric  sector   in the light--cone
gauge contains the two transverse space--time bosonic coordinates
$\bar X^{\mu}$ and 44 free fermions. The supersymmetric left
moving sector, in addition to the space--time bosons $X^{\mu}$ and
their fermionic superpartners $\psi^{\mu} $ includes also
 18 real free fermions $\chi^I,y^I,\omega^I$
($I=1,..., 6)$ among which supersymmetry is non--linearly
realized. The world--sheet supercurrent is
 \ba {T}_F =
\psi^{\mu}\partial{X}_{\mu}+\sum_I\chi^I\,y^I\omega^I
 \ea
 Then,
the theory is invariant under infinitesimal
super--reparametrizations of the world--sheet as the conformal
anomaly cancels separately in each sector. Each world--sheet
fermion $f_i$ is allowed to pick up a phase $\alpha_{f_i}
{\varepsilon} (-1,1]$ under parallel transport around a
non--contractible loop of the world--sheet
\ba f_i \ra - e^{\imath \pi\alpha_{f_i}} f_i
\ea
A spin structure is then defined as a specific set of  phases for
all world--sheet fermions,
\ba \alpha &=&
[\alpha_{f_1^r},\alpha_{f_2^r},\dots,\alpha_{f_k^r};
            \alpha_{f_1^c},\alpha_{f_2^c},\dots,\alpha_{f_l^c}]
\label{veca} \ea where $r$ stands for real, $c$ for complex and
$k+2 l=64$. For real fermions the phases $\alpha_{f_i^r}$ have to
be integers while $\alpha_{\psi^{\mu}}$ is independent of the
space--time index $\mu$.

The partition function is then  defined as a sum over a set of
spin structures ($\Xi$)
\ba Z(\tau)&\propto&
  \sum_{\alpha,\beta\in\Xi} c\left(\begin{array}{c}
\alpha\\\beta\end{array}\right){\rm Z}
\left(\begin{array}{c}\alpha\\\beta\end{array}\right)\ ,
\ea
where ${\rm Z} \left(\alpha\atop \beta\right)$ is the contribution
of the sector with boundary conditions $\alpha,\beta$ along the
two non--contractible circles of the torus and $c\left(\alpha\atop
\beta\right)$ a phase related to the GSO projection. Both $\Xi$
and $c \left(\alpha\atop \beta\right)$ are subject to string
constraints which guarantee the consistency  of the theory.

 A string model in  the context of free fermionic
formulation of the four--dimensional superstring is constructed by
specifying  a set of $n$ basis  vectors \footnote{By ${\bf1}$ we
denote the vector where all fermions are periodic.}
$(b_0={\bf1},b_1,b_2,\dots,b_{n-1})$ of the form (\ref{veca})
(which generate $\Xi=\sum_i m_i b_i$) and a set of
$\frac{n(n-1)}{2}+1$ independent phases $c \left(b_i\atop
b_j\right)\,$. Once a consistent set of basis vectors and a choice
of projection coefficients is made, the gauge symmetry, the
massless spectrum and the superpotential of the theory are
completely determined. In particular, the massless states of a
certain sector $\alpha=(\alpha_L;\alpha_R)\in\Xi$ are obtained by
acting on the vacuum $\left|0\right>_{\alpha}$ with the bosonic
and fermionic mode operators.
 The massless states $(M_L^2=M_R^2=0)$ are found by the
  Virassoro mass formula
\ba M_L^2 &=&
-\frac 12 +\frac{\alpha_L\cdot\alpha_L}{8} + \sum_f frequencies
\nonumber\\ M_R^2 &=& -1 +\frac{\alpha_R\cdot\alpha_R}{8} + \sum_f
frequencies \nonumber
\ea
where the sum is over the oscillator
frequencies \ba \nu_f=\frac{1+\alpha_{f_i}}2+{\rm integer},\;\;
\nu_{f^*}=\frac{1-\alpha_{f_i}}2+{\rm integer} \ea
 The physical states  are  obtained after the application of the GSO
 projections demanding
\ba
\left(e^{\imath \pi b_iF_{\alpha}} -\delta_{\alpha} c^*
\left(\alpha\atop b_i\right)\right) \left|{\rm physical\
state}\right>_\alpha =0
\ea
where $\delta_{\alpha} = -1$ if $\psi^{\mu}$ is periodic in the
sector $\alpha$ and $\delta_{\alpha} = +1$ when $\psi^{\mu}$ is
antiperiodic. The operator $b_iF_{\alpha}$ is
\ba
b_iF_{\alpha} = \left(\sum_{f\in{\rm\scriptstyle left}}-
       \sum_{f\in{\rm\scriptstyle right}}\right) b_i(f)F_{\alpha}(f)
\ea
where $F_{\alpha}(f)$ is the fermion number operator counting each
fermion mode $f$ once and its complex conjugate $f^*$ minus once.
It should be remarked that in the sector where all the fermions
are antiperiodic there is always a state
$\left|\mu,\nu\right>=\Psi^{\mu}_{-1/2}(\overline{\partial
X})^{\nu}_{-1}\left|0\right>_0$ which survives all projections and
includes the graviton, the dilaton and the two--index
antisymmetric tensor.

The present string model is defined in terms of nine basis vectors
$\{S,b_1,b_2,b_3,b_4,b_5,b_6$, $\alpha$, $\zeta\}$ and a suitable
choice of the GSO projection coefficient matrix. The resulting
gauge group has a Pati--Salam ($ SU(4)\times SU(2)_L\times
SU(2)_R$) non--Abelian observable part, accompanied by four
Abelian ($U(1)$) factors and a hidden $SU(8)\times U(1)'$
symmetry. The nine basis vectors are the following
\ba
\begin{array}{llllllll}
\ze&=&\{& &;
&\bar{\Phi}^{1...8}\}\\
S&=&\{\psi^{\mu},&\chi^{1...6},&;&\}\\
b_1&=&\{\psi^{\mu},&\chi^{12},(y\bar y)^{3456}, &;&\bar
{\Psi}^{1...5},\bar{\eta}^{1} \}\\
b_2&=&\{\psi^{\mu},&\chi^{34},(y\bar y)^{12}, (\omega\bar
\omega)^{56}&;&\bar {\Psi}^{1...5},\bar{\eta}^{2}\}\\
b_3&=&\{\psi^{\mu},&\chi^{56}, (\omega\bar
\omega)^{1234}&;&\bar {\Psi}^{1...5},\bar{\eta}^{3}\}\\
b_4&=&\{\psi^{\mu},&\chi^{12}, (y\bar y)^{36}, (\omega\bar
\omega)^{45}&;&\bar {\Psi}^{1...5},\bar{\eta}^{1} \}\\
b_5&=&\{\psi^{\mu},&\chi^{34}, (y\bar y)^{26}, (\omega\bar
\omega)^{15}&;&\bar {\Psi}^{1...5},\bar{\eta}^{2}\}\\
b_6&=&\{&&;&\bar
{\Psi}^{1...5},\bar{\eta}^{123},\bar{\Phi}^{1...4}\}\\
\alpha&=&\{&(y\bar y)^{36}, (\omega\bar \omega)^{36}
\bar \omega^{24}&;&\bar
{\Psi}^{123},\bar{\eta}^{23},\bar{\Phi}^{45}\}
\end{array}
\label{strbas}
 \ea
The specific projection coefficients we are using are given in
terms  of the exponent coefficients $c_{ij}$ in the following
matrix
\ba
 c_{ij} =
\bordermatrix{
 & z & S & b_1 & b_2 & b_3 & b_4 & b_5 &  b_6  &\alpha\cr
   z  &  1    & 1    & 1    & 1    & 1    & 1    & 1    & 0    & 0    \cr
   S   & 1    & 0    & 0    & 0    & 0    & 0    & 0    & 1    & 1    \cr
  b_1   & 1    & 1    & 1    & 1    & 1    & 1    & 1    & 1    & 1    \cr
  b_2   & 1    & 1    & 1    & 1    & 1    & 1    & 1    & 1    & 1    \cr
  b_3   & 1    & 1    & 1    & 1    & 1    & 1    & 1    & 1    & 1    \cr
  b_4   & 1    & 1    & 1    & 1    & 1    & 1    & 1    & 1    & 1    \cr
  b_5   & 1    & 1    & 1    & 1    & 1    & 1    & 1    & 1    & 0    \cr
  b_6   & 0    & 1    & 0    & 0    & 0    & 0    & 0    & 0    & 0    \cr
  \alpha   & 1    & 1    & 1    & 1    & 1    & 1    & 0    & 1    & 1}
\label{pc}
 \ea
 where the relation of $c_{ij}$ with $c(b_i,b_j)$ is
$$c\left(\begin{array}{c} b_i\\b_j\end{array}\right) = e^{\imath
\pi c_{ij}}$$ All world--sheet fermions appearing in the vectors of
the above basis are assumed to have {\em periodic} boundary
conditions. Those not appearing in each vector are taken with {\em
antiperiodic} ones. We follow the standard notation used in
references \cite{aehn,ALR}. Thus,
$\psi^{\mu},\chi^{1...6},(y/\omega)^{1...6}$ are real left, $(\bar
y/\bar \omega)^{1...6}$ are real right, and
 $\bar {\Psi}^{1...5}\bar{\eta}^{123} \bar {\Phi}^{1...8}$
are complex right  world sheet fermions. In the above,
${\bf1}=b_1+b_2+b_3+\ze$ and the basis element $S$  plays the role
of the supersymmetry generator as it includes exactly eight left
movers. {}Further, $b_{1,2,3}$ elements reduce  the $N=4$
supersymmetries  into $N=1$, while the initial $O(44)$ symmetry of
the right--moving sector results to an observable $SO(10)\times
SO(6)$ gauge group at this stage. The $SO(10)$ part corresponds to
the five $\bar {\Psi}^{1...5}$ complex world sheet fermions while
all chiral families at this stage belong to the ${\bf16}$
representation of the $SO(10)$. Vectors $b_{4,5}$ reduce further
the symmetry of the left moving sector, while the introduction of
the vector $b_6$ deals with the hidden part of the symmetry.
Finally, the choice of the vector $\alpha$ determines the final
gauge symmetry (observable and hidden sector) of the model which
is
\ba
SU(4)\times O(4) \times {U(1)}^4\times \{U(1)'\times
SU(8)\}_{hidden}. \label{strsym}
\ea
The observable gauge group consists of the non--Abelian
$SO(6)\times O(4)$ symmetry which is isomorphic to the left--right
Pati--Salam symmetry \cite{PS}. There are also four
$U(1)_{i=1,...,4}$ factors related to $\bar\eta_{1,2,3}$-complex
and the $\bar\omega^{24}$-real pair of world--sheet fermions of
the right--moving sector. All the superfields of the observable
sector carry non--zero charges under these four $U(1)$ symmetries.
Therefore, the latter are expected to play a very important role
in the determination of the Yukawa couplings, the fermion mass
textures, $R$-parity violation and in general in all types of
Yukawa interactions of the model. We note here that the observable
fields do not carry charges under $U(1)'$.

The Abelian part of the group deserves a separate treatment since
this class of models in general possess $U(1)$ symmetries which
are anomalous. Indeed, while we find two of the $U(1)$ factors to
be traceless ${\rm Tr} {U(1)}_1={\rm Tr} {U(1)}' =0$, the other
three are traceful, with ${\rm Tr} {U(1)}_2={\rm Tr} {U(1)}_3=
{\rm Tr} {U(1)}_4=24$. However, the $U(1)$ charges can be defined
in such a way that only one combination is anomalous. Indeed, the
linear combination
 \ba
 {U(1)}_A&=&{U(1)}_2+{U(1)}_3+{U(1)}_4,
\ea
has $ {\rm Tr}{\tilde{U}(1)}_A=72$, while there are other three
combinations orthogonal to the one above, which are free of gauge
and  gravitational anomalies. These are,
 \ba
 {{\tilde U}(1)}_1&=&{U(1)}_1\nonumber\\
 {{\tilde U}(1)}_2&=&{U(1)}_2-{U(1)}_3\\
 {{\tilde U}(1)}_2&=&{U(1)}_2+{U(1)}_3 - 2 {U(1)}_4.
\nonumber
\ea

The choice of the projection coefficients shown in (\ref{pc}) has
led to the desired three generation model as well as some
refinements of the previously proposed theory \cite{ALR} which are
phenomenologically appealing and deserve some discussion. The most
important are, the new Yukawa couplings which give fermions
masses, the mirror symmetry of the massless spectrum and the
number of $SU(4)$ Higgs multiplets.

We start with the enumeration of representations candidates for
families and $SU(4)\times SU(2)_R$ breaking Higgs fields as they
appear in Appendix A. We first note that due to the presence of
the various $U(1)$-factors, there is an arbitrariness in the
embedding of the electromagnetic charge operator. We will discuss
this in detail in the end of this section, however, to start with,
we assume first the simplest case where $U(1)_{em}$ is defined in
the standard way, (as in the original PS--symmetry), i.e:
\ba
Y=\frac{1}{\sqrt{6}} T_4+\frac 12 T_L+ \frac 12 T_R\label{hyp}
\ea
where $T_6,T_L,T_R$ are the diagonal $SU(4), SU(2)_L$ and
$SU(2)_R$ generators respectively.  Then, the massless spectrum is
classified with respect to its group properties as follows:
\begin{itemize}
\item
There are three copies of $[({\bf4}, {\bf2}, {\bf1})+({\bf\bar 4},
{\bf1}, {\bf2})]$ representations, available to accommodate the
three generations.
\item
There is one $[({\bf\bar
4},{\bf1},{\bf2})+({\bf4},{\bf1},{\bf2})]$ pair which is
interpreted as the Higgs pair triggering the $SU(4)\times SU(2)_R$
breaking.
\item
One pair $[({\bf4},{\bf2},{\bf1})+({\bf\bar 4},{\bf2},{\bf1})]$,
(mirror to each other) replaces the second Higgs pair of the old
string version \cite{ALR}. Clearly, since there are no mirror
families observed in the light spectrum, they should decouple at
some high scale by forming a heavy mass state.
\item
There are a large number of singlet fields with zero electric
charge, while carrying quantum numbers under the four $U(1)$
factors. In the determination of the flat directions of the model,
their vevs have to be chosen in such a way so as to cancel the
$D$--term. These singlets couple to ordinary matter via
superpotential terms. When they develop vevs along certain flat
directions they may create a hierarchical fermion mass spectrum
through non--renormalizable couplings.
\item
There are eight hidden $SU(8)$-octet and octet--bar superfields
(charged under the $U(1)^4\times U(1)'$), which are also neutral
under the usual charge definition. They can also acquire non--zero
vevs  leading to additional mass terms for ordinary or exotic
matter fields.
\item
The exotic states of the model fall into two categories:\\ {\it
i)} There are two $SU(4)$ fourplets $H_4=({\bf4},{\bf1},{\bf1})$,
$\bar{H}_4= ({\bf\bar{4}},{\bf1},{\bf1})$. After the symmetry
breaking, they result to a ${\bf3}$ and ${\bf\bar{3}}$ pair with
charges $\pm \frac 16$ respectively and two singlets with charges
$\pm \frac 12$.\\ {\it ii)} The second kind of exotic fields
includes ten left--handed doublets $X_{iL}$ and an equal number of
right--handed ones $X_{iR}$ with charges $\pm \frac 12$. The
presence of exotic particles in the massless spectrum of the
theory is a generic feature of level $k=1$ string constructions.
Such states in general, are regarded as string models' ``Achilles
heel'' since it is likely that they remain in the light spectrum
down to the electroweak scale. There are mainly three solutions to
this problem : first, one may choose a suitable flat direction
where all of them become massive at a relatively large scale; as a
second possibility, one can properly modify the string boundary
conditions on the basis vectors, so that these states appear with
non--trivial transformation properties under a hidden non--Abelian
group. In this latter case the exotic states are confined at the
scale where the gauge coupling of the hidden group becomes strong
 \cite{LRT}. Clearly, for a given number of matter representations,
the higher the rank of the group, the larger the confinement
scale. As a third possibility, one may consider the modification
of the charge operator (\ref{hyp}) by the inclusion of additional
$U(1)$ factors. In the present construction, we will discuss in
some detail the last two possibilities. Later, we will give a
brief account for their possible relevance on recent cosmological
observations.
\end{itemize}

Let us point out here that the observable spectrum of the model
respects the symmetry discussed in Section 2 with respect to the
simultaneous interchanges ${\bf2}_L\lra {\bf2}_R$ and ${\bf4}\lra
{\bf\bar{4}}$. In particular, left handed generation superfields
are interchanged with right handed ones, while there is a similar
change of roles of the $SU(4)$ higgs and mirrors. We will also see
in the following sections that the tree--level and higher order
Yukawa interactions remain also unaltered under the above
interchanges. The above symmetry is broken however, by the vacuum
when consistent $F$-- and $D$--flatness solutions are found. We
will discuss this when the superpotential of the theory is
presented and the corresponding flatness conditions are derived in
the next sections.

\section{Symmetry breaking and hypercharge embedding in the string
model.}

 We will discuss now two related issues, the gauge symmetry breaking
pattern and the various consistent embeddings of the weak
hypercharge. After defining the consistent set of boundary
conditions (\ref{strbas}) described previously, one is left with
an effective theory based on the symmetry (\ref{strsym}) with the
following general characteristics. There is an effective
unification scale, namely the string scale $M_{string}$, where all
couplings -- up to threshold corrections -- attain a common value.
At this point one is left with an effective $N=1$ supergravity
theory while the gauge group structure is of the form
$G=\prod_nG_n$, containing an `observable' and a `hidden' part.
The two worlds are not completely decoupled. Hidden and observable
fields are charged under five Abelian factors. The first symmetry
breaking occurs when some of the singlet fields acquire vevs to
cancel the $D$--term. Depending of the choice of the singlet vevs
several (at most four) of the above $U(1)$'s break, the natural
breaking scale $M_A$ being of the order of the $D-$term, i.e,
\ba
M_A&\sim &\sqrt{\xi}\nonumber\\
   & = & \sqrt{\frac{Tr Q_A}{192}}\frac{g_{string}}{\pi}M_{Pl}
     =    \frac{\sqrt{3}}{2\pi}g_{string}M_{Pl}
\ea
where $Tr{Q_A}=72$ is the trace of the anomalous $U(1)_A$ and
$M_{Pl}\approx 4.2\times 10^{18}$GeV is the reduced Planck mass.
We note here that, if only the singlets were allowed to obtain a
non--vanishing vev, at most four of the $U(1)$'s break; no singlet
is charged under the Abelian symmetry $U(1)'$ which remains
unbroken at this stage. The observable part $SU(4)\times
SU(2)_L\times SU(2)_R$ has a rank larger than that of the MSSM
symmetry, which breaks down to the SM--gauge group at an
intermediate scale $M_{GUT}$, usually some two orders of magnitude
below the string scale.  The breaking occurs in the way described
in Section 2. The necessity of the $SU(4)\times SU(2)_R$ symmetry
breaking  together with the $D-$ and $F$--flatness conditions
require at least two of the $U(1)$ factors to break at a high
scale.

There is finally the hidden $SU(8)$ part. This symmetry stays
intact, as long as  the ${\bf 8}$ and $\overline{\bf 8}$ fields do
not acquire vevs. Note also that the octets are charged under
$U(1)'$. In many flat directions which will be discussed
subsequently, phenomenological requirements force some of the
octets to obtain vevs and the symmetry $SU(8)\times U(1)'$ breaks
to a smaller one. Now, a crucial observation (see the relevant
Table in the Appendix A) is that all ${\bf 8}$'s come with $U(1)'$
positive charge  ($+1)$ whilst all  $\overline{\bf{8}}$'s appear
with the opposite ($-1$) charge. It is easy to show then, that, no
matter how many of the available  octet fields receive a non--zero
vev, there is always  a $U(1)''$ unbroken which is a linear
combination of the $U(1)'$ and one of the generators of the
$SU(8)$. Therefore, the hidden matter conserves a new $U(1)''$
symmetry which stays unbroken down to low energies. Its
cosmological implications will be discussed in a subsequent
section.

We turn now our discussion to the hypercharge embedding. As
mentioned above, due to the appearance of extra $U(1)$--factors,
the hypercharge generator is not uniquely determined. It can be
any linear combination of the $U(1)_{B-L}$ the five available
$U(1)$'s of the model and possible unbroken generators of the
hidden gauge symmetry, provided the minimal supersymmetric low
energy particle spectrum is generated. The standard weak
hypercharge assignment -- as this is defined in the original
PS--symmetry -- does not involve any of the surplus $U(1)$ factors
discussed above. It is solely determined in the usual sense from
the diagonal generators of the PS--symmetry as in (\ref{hyp}).
Under this definition in addition to the representations
accommodating the MSSM--fields, the states found in
$({\bf4},{\bf1},{\bf1})+({\bf\bar 4},{\bf1},{\bf1})$ and
$({\bf1},{\bf1},{\bf2})+({\bf1},{\bf2},{\bf1})$ representations
obtain the exotic charges discussed above, while they are rather
unusual in the grand unified models.

We will discuss here in some detail another possible definition of
the hypercharge operator which is obtained by including the
$U(1)'$-generator:
\ba Y'=\frac{1}{\sqrt{6}} T_4+\frac 12 T_L+
\frac 12 T_R- \frac{\omega}{2} Q' \label{hypr}
\ea
where $Q'$ is related to the $U(1)'$ charge of the particular
massless state and $\omega$ is the appropriate normalization
constant. Choosing for example $\omega =1$, all extra doublets
$X_{L,R}$ obtain integral charges ($\pm 1, 0$). On the other hand,
this new embedding leads to the normalization of the hypercharge
generator
\ba k=\frac{5+ 12\omega^2}{3}. \ea The value of the weak mixing
angle at $M_{\rm string}$ is $\sin^2\theta_W(M_{\rm string}) =
\frac{1}{1+k}$. Its values for the two lower $\omega$'s are given
in  Table 1.
\begin{table}
\begin{center}
\begin{tabular}{|c|c|} \hline
  $\omega$ & $\sin^2\theta_W$
\\ \hline
  0 &$ \frac{3}{8}\mathstrut$\\\hline
  1 & $ \frac{3}{22}$\mathstrut\\
\hline
\end{tabular}
\end{center}
\label{tab:omega}
\caption{{The values of the weak mixing angle at the Unification
scale, for two definitions of the weak hypercharge. In the second
case, an additional
$U(1)$-factor is assumed.}}
\end{table}
{}For $\omega =0$, we obtain the standard GUT  $\sin^2\theta_W$
prediction but the exotic states have fractional charges, whereas
for  $\omega =1$ the $X_{L,R}$ doublets as well as the
$({\bf4},{\bf1},{\bf1})$ and $({\bf\bar 4},{\bf1},{\bf1})$
representations, obtain charges like those of the ordinary down
quarks and leptons. It should not escape our attention that in
this new hypercharge definition $Y'$ the octet fields now appear
with fractional charges $\pm 1/2$. This is not however a real
problem. The coupling of the $SU(8)$ group becomes strong at a
high scale, leading to a confinement (in close analogy with QCD),
and forcing the octets to form bound states with the corresponding
octet--bar fields. We should note here that this situation opens
up the possibility of giving vevs to these condensates at a
smaller scale and create new mass terms for the ordinary matter
through their superpotential couplings. The new hypercharge
definition predicts a low value for the weak mixing angle which is
essentially the value obtained in a Kac--Moody level $k=2$ string
construction~\cite{schell}. Starting however, from such a small
initial value for $\sin^2\theta_W$ at $M_{\rm string}$, there is
no obvious way how the larger low energy value can be obtained in
this case.

Let us close this section with a short comment on one more
possibility of symmetry breaking. One can give vevs directly to
the exotic $SU(2)_R$-doublet fields $X_{iR}$. (Both their
components are charged ($Q_{em}=\pm 1/2$) under the standard
hypercharge assignment (\ref{hyp})). The vev should be taken along
the neutral direction defined by the proper linear combination.
This will essentially lead to a definition of the hypercharge
as in the case of $\omega =1$. However, this approach has the
advantage that the small initial value of $\sin^2\theta_W$ is defined
now at the $SU(2)_R$-breaking scale which can be taken to be much
lower that the string scale. This case requires a separate treatment
since there are new fields entering the flatness conditions while
new mass terms appear in the superpotential. Moreover, exotic
doublets now look like the ordinary electron doublets while
far reaching phenomenological implications appear.

\section{The Superpotential of the string version.}

We proceed now to the calculation of the perturbative
superpotential. Clearly, the number of fields in the string
version is significantly larger than those of its surrogate GUT
discussed in Section 2. In fact, in the model of Section 2, the
construction of the superpotential was rather easy since only
gauge symmetries had to be respected. Here, however, not all gauge
invariant terms are allowed; additional restrictions from
world--sheet symmetries have to be taken into account since they
eliminate a large portion of the gauge invariant superpotential
terms. A short description of the calculation \cite{RT,KLN} of the
renormalizable as well as the non--renormalizable superpotential
terms is given in the Appendix B.

The tree--level  superpotential is \ba \frac{W_3}{g\sqrt{2}}
&=&\nonumber\\&&\mbox{}
   {\overline{F}_{5R}} {{F}_{4L}} {{h}_{4}}+
   {\overline{F}_{3R}} {{F}_{3L}} {{h}_{2}}+
   \frac{1}{\sqrt{2}}{\overline{F}_{5L}} {{F}_{4L}} {{\zeta}_{2}}+
   \frac{1}{\sqrt{2}}{\overline{F}_{5R}} {{F}_{4R}} {\overline{\zeta}_{3}}+
   \nonumber\\&&\mbox{}
   +
   {\overline{F}_{5L}} {\overline{F}_{5L}} {{D}_{4}}+
   {{F}_{4L}} {{F}_{4L}} {{D}_{1}}+
   {{F}_{2L}} {{F}_{2L}} {{D}_{2}}+
    {{F}_{1L}} {{F}_{1L}} {{D}_{1}}+
   {{F}_{1L}} {{\overline{H}_4}} {{X}_{7L}}+
     \nonumber\\&&\mbox{}
     +
   {\overline{F}_{1R}} {\overline{F}_{1R}} {{D}_{1}}+
    {{F}_{4R}} {{F}_{4R}} {{D}_{3}}+
   {\overline{F}_{2R}} {\overline{F}_{2R}} {{D}_{2}}+
   {\overline{F}_{5R}} {\overline{F}_{5R}} {{D}_{2}}+
     {\overline{F}_{2R}} {{H}_4} {{X}_{3R}} +
    \nonumber\\&&\mbox{} +
   \frac{1}{2}{{\Phi}_{2}}\left( {{\zeta}_{i}} {\overline{\zeta}_{i}}+
   {{\xi}_{i}} {\overline{\xi}_{i}}+
    {{h}_{3}} {{h}_{4}}+
    {{\overline{H}_4}} {{H}_4}\right)
     \nonumber\\&&\mbox{} +
   {{\Phi}_{4}} \left({{\zeta}_{1}} {\overline{\zeta}_{3}}+
   {{\zeta}_{3}} {\overline{\zeta}_{1}}\right)+
   {{\Phi}_{5}} \left({{\zeta}_{2}} {\overline{\zeta}_{4}}+
   {{\zeta}_{4}} {\overline{\zeta}_{2}} \right) \nonumber\\&&\mbox{} +
   {{D}_{1}} {{D}_{2}} {{\Phi}_{12}}+
   {{D}_{1}} {{D}_{4}} {{\Phi}_{12}^{-}}+
   {{D}_{2}} {{D}_{3}} {\overline{\Phi}_{12}^{-}}+
   {{D}_{3}} {{D}_{4}} {\overline{\Phi}_{12}}\nonumber\\&&\mbox{} +
   {{h}_{2}} {\overline{\xi}_{1}} {{h}_{4}}+
   {{h}_{2}} {\overline{\xi}_{4}} {{h}_{3}}+
   {{h}_{2}} {{X}_{10L}} {{X}_{10R}}+
   {{h}_{1}} {{\xi}_{1}} {{h}_{3}}+
   {{h}_{1}} {{\xi}_{4}} {{h}_{4}}+
   {{h}_{1}} {{X}_{9L}} {{X}_{9R}}\nonumber\\&&\mbox{} +
   {\overline{\Phi}_{12}} \left({{\xi}_{4}} {\overline{\xi}_{1}}+
   {{h}_{3}} {{h}_{3}}+
    {{Z}_{3}} {\overline{Z}_{3}}\right)+
   {\overline{\Phi}_{12}^{-}}\left( {\overline{\zeta}_{i}}
   {\overline{\zeta}_{i}}+
   {{\xi}_{3}} {\overline{\xi}_{2}}+
    {{X}_{9R}} {{X}_{10R}}\right)\nonumber\\&&\mbox{} +
   {{\Phi}_{12}^{-}} \left({{\zeta}_{i}} {{\zeta}_{i}}+
   {{\xi}_{2}} {\overline{\xi}_{3}}+
   {{X}_{9L}} {{X}_{10L}}\right)+
   {{\Phi}_{12}} {{\xi}_{1}} {\overline{\xi}_{4}}+
   {{\Phi}_{12}} {{h}_{4}} {{h}_{4}}+\nonumber\\&&\mbox{} +
   \frac{1}{\sqrt{2}}\left({{\zeta}_{1}} {{X}_{1R}} {{X}_{6R}}+
   {{\zeta}_{3}} {{Z}_{4}} {\overline{Z}_{5}}+
   {{\zeta}_{4}} {{X}_{2R}} {{X}_{5R}}+
   {\overline{\zeta}_{2}} {{Z}_{5}} {\overline{Z}_{4}}+
   {\overline{\zeta}_{4}} {{X}_{1L}} {{X}_{6L}}+
   {\overline{\zeta}_{1}} {{X}_{2L}} {{X}_{5L}}\right)
   \nonumber\\&&\mbox{} +
   {{\xi}_{1}} {{Z}_{5}} {\overline{Z}_{5}}+
   {{\xi}_{2}} {{X}_{2R}} {{X}_{6R}}+
   {\overline{\xi}_{2}} {{X}_{1L}} {{X}_{5L}}+
   {{h}_{3}} {{X}_{2R}} {{X}_{5L}}+
   {{h}_{4}} {{X}_{1L}} {{X}_{6R} }\ .
   \label{sup}
 \ea
 {}The fourth order superpotential terms are
 \ba
 w_4&=&
   {\overline{F}_{5L}} {{F}_{4L}} {{X}_{1L}} {{X}_{6L}}+
   {\overline{F}_{5L}} {{F}_{3L}} {{Z}_{3}} {\overline{Z}_{4}}+
   {\overline{F}_{5L}} {{F}_{1L}} {{X}_{4L}} {{X}_{6L}}+
   {\overline{F}_{5R}} {{F}_{4R}} {{X}_{1R}}
   {{X}_{6R}}+\nonumber\\
   &&\mbox{}
   {{F}_{4R}} {\overline{F}_{3R}} {\overline{Z}_{3}} {{Z}_{4}}+
   {{F}_{4R}} {\overline{F}_{2R}} {{X}_{1R}} {{X}_{8R}}+
   {\overline{F}_{2R}} {{F}_{2L}} {\overline{\zeta}_{4}} {{h}_{4}}+
   {\overline{F}_{1R}} {{F}_{1L}} {{\zeta}_{1}} {{h}_{4}}+
   \nonumber\\
   &&\mbox{}
   {\overline{\zeta}_{1}} {{\xi}_{1}} {\overline{Z}_{1}} {{Z}_{1}}+
   {{\zeta}_{2}} {{h}_{4}} {{X}_{8L}} {{X}_{8R}}+
   {{\zeta}_{2}} {{\xi}_{3}} {{X}_{7R}} {{X}_{8R}}+
   {{\zeta}_{2}} {\overline{\xi}_{2}} {{X}_{7L}} {{X}_{8L}}+
   \nonumber\\
   &&\mbox{}
   {\overline{\zeta}_{2}} {{h}_{3}} {{X}_{7L}} {{X}_{7R}}+
   {\overline{\zeta}_{2}} {{\xi}_{2}} {{X}_{7R}} {{X}_{8R}}+
   {\overline{\zeta}_{2}} {\overline{\xi}_{3}} {{X}_{7L}} {{X}_{8L}}+
   {{\zeta}_{3}} {{h}_{3}} {{X}_{3L}} {{X}_{3R}}+
   \nonumber\\
   &&\mbox{}
   {{\zeta}_{3}} {{\xi}_{3}} {{X}_{3R}} {{X}_{4R}}+
   {{\zeta}_{3}} {\overline{\xi}_{2}} {{X}_{3L}} {{X}_{4L}}+
   {\overline{\zeta}_{3}} {{h}_{4}} {{X}_{4L}} {{X}_{4R}}+
   {\overline{\zeta}_{3}} {{\xi}_{2}} {{X}_{3R}} {{X}_{4R}}+
   \nonumber\\
   &&\mbox{}
   {\overline{\zeta}_{3}} {\overline{\xi}_{3}} {{X}_{3L}} {{X}_{4L}}+
   {{\zeta}_{4}} {{\xi}_{1}} {\overline{Z}_{2}} {{Z}_{2}}+
   {{Z}_{1}} {\overline{Z}_{4}} {{X}_{3R}} {{X}_{5R}}+
   {\overline{Z}_{2}} {{Z}_{4}} {{X}_{2L}} {{X}_{7L}}+
   \nonumber\\
   &&\mbox{}
   {{Z}_{4}} {\overline{Z}_{5}} {{X}_{2L}} {{X}_{5L}}+
   {{Z}_{5}} {\overline{Z}_{4}} {{X}_{2R}} {{X}_{5R}}+
   {{X}_{3L}} {{X}_{4R}} {{X}_{10L}} {{X}_{9R}}+
   {{X}_{4L}} {{X}_{3R}} {{X}_{9L}} {{X}_{10R}}+
   \nonumber\\
   &&\mbox{}
   {{X}_{1L}} {{X}_{2R}} {{X}_{9L}} {{X}_{10R}}+
   {{X}_{2L}} {{X}_{1R}} {{X}_{9L}} {{X}_{10R}}+
   {{X}_{5L}} {{X}_{6R}} {{X}_{10L}} {{X}_{9R}}+
     \nonumber\\
   &&\mbox{}
   {{X}_{6L}} {{X}_{5R}} {{X}_{10L}} {{X}_{9R}}+
   {{X}_{7L}} {{X}_{8R}} {{X}_{10L}} {{X}_{9R}}+
   {{X}_{8L}} {{X}_{7R}} {{X}_{9L}} {{X}_{10R}} \ ,
   \label{4nr}
\ea
where in each terms an  ${\cal O}(1)g/M_{Pl}$ multiplicative factor.
 Higher order terms up to sixth order have been also calculated
and are presented in the Appendix B.

Having obtained the spectrum of the model, as well as the
available superpotential terms, we need to determine the vacuum of
our theory, by making an appropriate choice of the vacuum
expectation values of the Higgs fields (fourplets, bidoublets and
a sufficient number of singlets)  and possibly some of the hidden
$SU(8)$ multiplets. All these choices should be consistent with
the $D-$ and $F-$ flatness conditions. A complete account of all
possible solutions of these conditions will be given subsequently,
however, not all of those solutions are satisfactory from the
phenomenological point of view. A final conclusion about the
viability of a certain flat direction however cannot be drawn
before adequately high order NR--terms are taken into account.
There are two main reasons for this: first, it is possible that a
particular viable  flat direction at a certain order, is destroyed
when  higher order NR--terms are included in the calculation.
Second, even if a phenomenologically promising flat direction can
be proven to persist at higher orders, it is possible that the new
NR superpotential terms create undesired mass terms. For example,
a usual phenomenon is that they fill in many entries in the Higgs
mass matrix, so it is possible that there is no massless higgs
left to break the symmetry. As a consequence, one may have further
constraints on the particular flat direction by forcing some
additional fields to obtain a zero vev. This will be discussed in
a subsequent section.

{}From the above remarks, it is evident that the right choice of
the vacuum of the model is not an easy task. In the next section
our endeavor will be concentrated in the classification of all
flat directions and their relevance to the low energy
phenomenological expectations. It is useful therefore, in order to
pin down the few promising vacua from the hundreds of available
solutions, to summarize the basic observations which will help us
to complete this task. This will enable us to determine the right
flat direction and choose those singlet (and possibly hidden)
field vevs that guarantee a successful description of the low
energy phenomenological theory.

\begin{itemize}
\item
 We start with the Higgs mechanism; we first observe that there is
only one pair of Higgs fields available to break $SU(4)$ symmetry,
namely the fourplet $F_{4R}$ and in general  one linear
combination of the fields $\bar{F}_{1,2,3,5}$. Thus, in order to
keep $F_{4R}$ massless and prevent a mass term through the
tree--level coupling $F_{4R}\bar{F}_{5R}\bar\ze_3$, we need to
impose $\langle\bar\ze_3 \rangle = 0$.

\item
In addition to the three generations expected to appear at low
energies, the model predicts also the existence of one additional
$({\bf4},{\bf2},{\bf1})$ representation plus its mirror
$\bar{F}_{5L} = (\bar{4},2,1)$. Since no mirror families appear in
the low energy spectrum, we need a mass term of the form
$\langle\chi_i\rangle F_{iL} \bar{F}_{5L}$ (where $\chi_i$ some of
the singlets with non--zero vevs) to give a heavy mass to the
mirror $\bar{F}_{5L}$. A candidate term could be
$\ze_2\bar{F}_{5L}F_{4L}$ which exists already at the tree--level.
 At fourth order
there is also the term $ \bar{F}_{5L}F_{3L} Z_3\bar{Z}_5$.
Thus, up to fourth order, we obtain
\ba
 \sum_i \langle\chi_i\rangle F_{iL} \bar{F}_{5L}=
    (\ze_2F_{4L}+ F_{3L} Z_3\bar{Z}_4
       +\cdots )
    \bar{F}_{5L}
\label{mir1} \ea where $\{\cdots\} $ stand for possible higher
order NR--terms involving fields that may acquire vevs. Clearly,
if we wish to make the mirror multiplets heavy with superpotential
terms up to fourth order, we should demand from flatness
conditions either $\ze_2\not=0$, or $Z_3\bar{Z}_5\not=  0$.
Solutions with higher order NR--terms are also possible as it will
be clear later.

\item
A number of sextet fields, $D_i$, $i=1,...,4$ containing colour
 triplets as well as triplets surviving the Higgs mechanism
   appear also in the spectrum. In order to avoid possible proton
    decay problems we need
also mass terms for those coloured fields. As far as the sextet
fields  are concerned, the sextet matrix at tree--level is
\ba
{{D}_{1}} {{D}_{2}} {{\Phi}_{12}}+
   {{D}_{1}} {{D}_{4}} {{\Phi}_{12}^{-}}+
   {{D}_{2}} {{D}_{3}} {\overline{\Phi}_{12}^{-}}+
   {{D}_{3}} {{D}_{4}}\bar\Phi_{12}
\ea
Their eigenmasses in terms of the scalar components of the
singlet fields are
\ba
m_{D_i}&=&\pm\frac 12
          \left(\Sigma_{\Phi^2}\pm\sqrt{(\Sigma_{\Phi^2})^2
     -(\Phi_{12}\bar\Phi_{12}-\Phi_{12}^-\bar\Phi_{12}^-)^2}
          \right)
          \label{mcs}
\ea
with
\ba
\Sigma_{\Phi^2}&=& {\Phi}_{12}^2+\bar\Phi_{12}^2
         +{{\Phi}_{12}^{-}}^2+{\overline{\Phi}_{12}^{-}}^2
\ea
The above eigenvalues are all non--zero whenever the condition
$$\Phi_{12}\bar\Phi_{12}-\Phi_{12}^-\bar\Phi_{12}^-\not=0$$ is
fulfilled. Therefore, a satisfactory flat direction should  keep
the appropriate  singlets  with non--zero vevs. It will be clear
later that in most of the phenomenologically viable string vacua,
higher order NR--contributions will prove necessary to make all
sextet fields massive.

In forming the mass matrices for triplets, we should also take
into account the fact that there are also coloured triplets in the
Higgs pair $({\bf\bar 4},{\bf1},{\bf2}) +({\bf4},{\bf1},{\bf2})$.
Thus, recalling in mind the sextet decomposition $ D_4\ra
{D}_4^{\bar 3} +{D}_4^{3}$, the term $F_{4R}F_{4R} D_4$ gives a
heavy mass to $\bar d^c_{4R}{D}_4^{\bar 3}$  and the terms
\ba
 \bar F_{5R}   \bar F_{5R}    D_2
   +           \bar F_{2R}   \bar F_{2R}   D_2
   +           \bar  F_{1R}  \bar  F_{1R}   D_1
\nonumber
\ea
make another  $\bar 3 - 3$ combination massive. This linear
combination depends on the choice of the fields which are going
to accommodate the families and higgses. This will be precisely
determined as long as a specific flat direction is chosen.

\item
 There is a new interesting feature of this version;
There are two candidate terms at fourth order to provide
 masses to the  second generation
\ba \bar F_{1R}F_{1L}{h}_{4}\zeta_1+\bar
F_{2R}F_{2L}{h}_4\bar\zeta_4 \ea They are expected to be of the
right order, provided that at least one of the singlets
$\ze_1,\bar\ze_4$ gets a non--zero vev.
\end{itemize}

After this preliminary analysis, we are ready now to explore other
important aspects of the model.  In the next section we will find
all tree--level and higher order solutions to the flatness
conditions which determine  the consistent string vacua.

\section{ The solutions of the F-- and D--flatness conditions}

One of the main concerns in constructing effective supersymmetric
models from superstrings, is to find the flat directions along
which the scalar potential vanishes. String models in general
contain several flat directions which are lifted by higher order
corrections to the superpotential and supersymmetry breaking
effects. The latter set also the scale of the scalar potential.
Another interesting fact in string model building concerning these
flat directions, is the existence of a $D$--term contribution
\cite{FI,DSW}. As has been discussed in the previous section,
there is a linear combination of the four surplus $U(1)$ factors
accompanying the observable gauge group of the model, which
remains anomalous. The standard anomaly cancellation mechanism
\cite{DSW} results to a shift of the vacuum where several scalar
components of the singlet (and possibly hidden) superfields
develop non--zero vevs. Their magnitude is determined by the
solution(s) of the complete system of the $F$- and $D$-flatness
constraints.

{\bf Derivation of the flatness constraints}

The $F$-flatness equations are easily derived from the
superpotential. They are the set of equations resulting from the
differentiation of ${\cal W}$ with respect to the fields of the
massless spectrum $f_i$, \ba \frac{\partial }{\partial f_i} {\cal
W}= 0\nonumber \ea In this paper we will mainly concentrate on an
analysis of the flatness conditions involving fields only from the
observable sector. For completeness, we also give in the Appendix
B the relevant contributions to the flatness conditions taking
into account hidden field vevs as well as higher order corrections
from NR--terms.

 Taking the derivatives of the renormalizable superpotential
${\cal W}$ with respect to the observable fields, we obtain
 \ba
\Phi_2: &  \zeta_i     \bar \zeta_{i}
                       = -\xi_i    \bar\xi_{i}
\label{ 1}\\
\Phi_4: &  \zeta_1  \bar\zeta_{3}
                   = -\bar\zeta_{1} \zeta_3
\label{ 2}\\
\Phi_5: & \zeta_2 \bar\zeta_{4}
                   = -\bar\zeta_{2} \zeta_4
\label{ 3}\\
\Phi_{12}: &
                        \xi_1 \bar \xi_{4}=0
\label{ 4}\\
\bar\Phi_{12}: &
                     \xi_4   \bar \xi_{1}
          = 0
\label{ 5}\\
\Phi_{12}^-: &  \zeta_i \zeta_i
                   = - \xi_2   \bar \xi_{3}
\label{ 6}\\
\bar\Phi_{12}^-: & \bar\zeta_{i}\bar\zeta_{i}
                         =- \xi_3\bar \xi_{2}
\label{ 7}\\
\xi_1 : & \phi_2\bar\xi_1
                     = - \Phi_{12}\bar\xi_4
\label{ 8}\\
\bar\xi_1 : &  \phi_2 \xi_1
                      = - \bar\Phi_{12}\xi_4
\label{ 9}\\
\xi_2:& \phi_2\bar\xi_2 = -  \Phi_{12}^-\bar\xi_3
\label{ 10}\\
\bar\xi_2:& \phi_2\xi_2 = - \bar\Phi_{12}^-\xi_3
\label{ 11}\\
\xi_3:& \phi_2\bar\xi_3 = - \bar\Phi_{12}^-\bar\xi_2
\label{ 12}\\
\bar\xi_3:&\phi_2\xi_3
                      = - \Phi_{12}^-\xi_2
\label{ 13}\\
\xi_4:& \phi_2\bar\xi_4
         = - \bar\Phi_{12}\bar\xi_1
\label{ 14}\\
\bar\xi_4:& \phi_2 \xi_4
                 = - \Phi_{12} \xi_1
\label{ 15}\\
\zeta_1:& \phi_2\bar\ze_1+\Phi_4\bar\ze_3 + 2\Phi_{12}^-\ze_1 = 0
\label{ 16}\\
\bar\ze_1:& \phi_2\ze_1 +\Phi_4\ze_3 + 2\bar\Phi_{12}^-\bar\ze_1
 = 0
\label{ 17} \\
\ze_2: & \phi_2\bar\ze_2 +\Phi_5\bar\ze_4 + 2\Phi_{12}^-\ze_2
                        +\bar{F}_{5L}F_{4L}/\sqrt{2}= 0
\label{ 18} \\
\bar\ze_2: & \phi_2\ze_2 +\Phi_5\ze_4 + 2\bar\Phi_{12}^-\bar\ze_2
 = 0
\label{ 19}\\
\ze_3:& \phi_2\bar\ze_3 +\Phi_4 \bar\ze_1 + 2\Phi_{12}^-\ze_3
 = 0
\label{ 20}\\
\bar\ze_3 :&  \phi_2\ze_3 +\Phi_4\ze_1 + 2\bar\Phi_{12}^-\bar\ze_3
           + \bar F_{5R}F_{4R}/\sqrt{2} = 0
\label{ 21}\\
\ze_4: & \phi_2\bar\ze_4 + \Phi_5\bar\ze_2 + 2\Phi_{12}^-\ze_4
 = 0
\label{ 22} \\
\bar\ze_4: & \phi_2\ze_4 +\Phi_5\ze_2 + 2 \bar\Phi_{12}^-\bar\ze_4 = 0
\label{ 23}
\ea
On the left of the above equations we show the field with respect
to which the superpotential is differentiated . In equations
(\ref{ 18},\ref{ 21}) both $SU(2)_L$ and $SU(2)_R$ fourplet fields
have  been included to exhibit the invariance of the equations
under a straightforward generalization of the transformations
(\ref{msym}). Indeed, it can be observed now that the
$F$--flatness equations  as well as the Yukawa interactions remain
unaltered under the interchanges mentioned in  previous sections.
In particular, when ${\bf\bar{4}}$ and ${\bf2}_R$ are interchanged
with $\bf4$ and ${\bf2}_L$ respectively, it can be seen  that the
superpotential remains invariant under the following renaming of
the fields
\ba
F_{5},\bar{F}_5 \lra F_4, \bar{F}_4& F_{1},\bar{F}_1\lra F_2,
\bar{F}_2 & D_{1},D_{3}\lra D_{2},D_{4}\\
\ze_1,\ze_2 \lra \bar\ze_4,\bar\ze_3&\bar\ze_1, \bar\ze_2 \lra
\ze_4,\ze_3 &\Phi_1,\Phi_4 \lra \Phi_3,\Phi_5\\ \xi_2, \xi_3\lra
\bar\xi_2,\bar\xi_3&\Phi_{12}^-\lra \bar\Phi_{12}^-&
\Phi_2\lra \Phi_2
\ea
This symmetry is also preserved by higher order Yukawa terms as
can be easily checked from the terms presented in the Appendix.
Nevertheless, the $D$--equations are not invariant. Clearly, any
solution of them defines a vacuum which does not preserve the
symmetry. In the subsequent, we make a definite choice with regard
to this symmetry putting all $\langle F_{iL}\rangle = 0$.

The $D$--flatness equations for anomalous or non--anomalous $U(1)$
factors with hidden field contributions are also derived  in the
Appendix C. In the absence of non--zero hidden field vevs, they
can be written in the following compact form \ba ({\cal
D}_1):\;&f_3& = x_2 + x_3 \label{d1}\\ ({\cal D}_2):\;&f_4-f_1 &=
x_2 +\frac{\ze}2-\bar\phi \label{d2}\\ ({\cal D}_3):\;&\sum_i x_i
&=-\frac{\xi}3 \label{d3}\\ ({\cal D}_4):\;&\phi &=x_1
+\frac{\xi}2 \label{d4}\\ ({\cal D}_5):\;&f_4-f_1 &= f_2+f_3+f_5
\label{d5} \ea where, we have defined, \ba x_i =
|\bar\xi_i|^2-|\xi_i|^2;\; &\ze =
\sum_i(|\bar\ze_i|^2-|\ze_i|^2)\\ \phi =
|\bar\Phi_{12}|^2-|\Phi_{12}|^2;\; &\bar\phi =
|\bar\Phi_{12}^-|^2-|\Phi_{12}^-|^2\\ f_4  = \frac 12
|{F}_{4R}|^2;\; &f_i=  \frac 12 |\bar{F}_{iR}|^2, i=1,2,3,5 \ea

{\bf Tree and higher level solutions of $F$-- and $D$--flatness
constraints}

A consistent phenomenological analysis of the model requires a
complete knowledge of all vacua, therefore, a systematic approach
to classify all $D$-- and $F$--flat directions is needed.  When
this is done, we will be able to know  which fields acquire
non--zero vevs in any specific flat direction{\footnote{{}For
similar systematic analyzes in other models, see
\cite{RTA,art,cceel,pr}.}. These vevs will determine completely the
masses of fermion and scalar fields through their superpotential
couplings.

In the remaining of this section we present a systematic analysis
of the above constraints, taking into account basic
phenomenological requirements. This will limit considerably the
number of possible solutions. Thus, for example, as already has
been pointed out, it is necessary to impose the constraint
$\langle\bar\ze_3\rangle = 0$, in order to prevent a mass term for
the Higgs field $F_{4R}$ at tree--level. This, by no means ensures
the existence of a consistent solution. We postpone the complete
presentation after we obtain the set of mathematically consistent
cases. In the present paper we restrict the analysis of the
flatness conditions in the case where only observable fields
acquire non--zero vevs. Solutions with hidden field vevs are much
more involved and may result to interesting new vacua. Although a
detailed investigation of the latter will not considered here, a
brief discussion of their role is given later in this paper.

{}We find it convenient to start our analysis from the $F$--flatness
conditions  (\ref{ 3},\ref{ 4}) which imply four distinct cases:
$$
\begin{array}{ll}
(i)\ \ \   \bar\xi_1=\xi_1 = 0\ ,\ \xi_4\ne0;\;\; & (ii)\; \xi_1 =
 \xi_4 = 0;\\
(iii)\; \xi_4 = \bar\xi_1 = 0;\;\; &(iv)\; \bar\xi_4 = \bar\xi_1 =
 0\ , \xi_1,\xi_4\ne0\ .
\end{array}
$$
{}From the above cases, only $(iii,iv)$ have consistent solutions.
Let's first explain why cases $(i)$ and $(ii)$ are rejected.

\underline{Cases $(i)$ and $(ii)$}\\ {}From Eqs. (\ref{ 14},\ref{
15}) we deduce $\phi_2=0$, while $\xi_4\not=0$ in Eq. (\ref{ 9})
imposes $\bar\Phi_{12}=0$. This however leads to inconsistency
with equation ${\cal D}_4$, since the left side of the equation is
negative while the right side is positive and non--zero. Similarly
case $(ii)$ where $\xi_1 = \xi_4 = 0$, is not soluble as can be
easily seen from the equation (${\cal D}_3$)+(${\cal D}_1$). We
consider the two remaining cases $(iii)$ and $(iv)$separately.

\underline{Case $(iii)$}\\ {}From Eq. (\ref{ 9}), we find
$\phi_2=0$. Further, ${\cal D}_4$--flatness tells us that $\xi_1$
and $\bar\Phi_{12}$ cannot be simultaneously zero. Then, combining
this with  conditions (\ref{ 13}) and (\ref{ 14}) we conclude that
\ba \Phi_{12}=0,\; \bar\xi_4=\xi_4 =0,\; \xi_1\not=0, \; \phi_2=0
\ea while $\bar\xi_1\cdot\bar\Phi_{12} = 0$.

Proceeding further, we classify all solutions in this case
according to their number of free parameters and fields with zero
vevs. At the tree--level, there are 17 solutions consistent with
the $F$-- and $D$--flatness conditions. These are cases 1--17 of
Table 9 of the Appendix D. Several of these flat directions are
lifted when higher order NR--terms are included. On the other
hand, other tree--level flat directions  remain flat when
additional constraints have are imposed on the field vevs. There
are cases where a single tree--level flat direction results to
more than one distinct cases at a higher level since the solution
of the constraints may be satisfied for various choices of field
vevs.

When NR contributions to flatness conditions up $6^{th}$ order are
taken into account, the above tree--level solutions reduce to the
first thirteen cases presented in  Table 2.  The first column
numbers the solutions, while the last one denotes the number of
 free (complex) parameters left. The  five columns  in the middle
show the fields with zero vevs, where for presentation purposes
abbreviations in the field notation have been used. Thus, in the
second row, the numbers $12, \overline{12}, 12^-,
\overline{12}^-$,  denote the fields $\Phi_{12},
\overline{\Phi}_{12}$,$\Phi_{12}^- , \overline{\Phi}_{12}^-$, and
so on.  The fields which are forced to obtain zero vevs due to
higher order NR--contributions in the Yukawa potential, are
included in curly brackets. Thus, for example, in the  fourth
column of the first case, the symbol $\{\bar 2\}$ means that
$\langle\bar{\xi}_2\rangle $ has a zero vev due to the inclusion
of NR--terms. Further, for the same reason in the fifth column  we
also use the notation $\{\bar 1\},\{\bar 2\}$ which should be
translated to  the conditions $\bar{F}_{1R} = \bar{F}_{2R} = 0$
imposed by  NR--terms. In this notation, one can see the effect of
NR--terms in the tree--level solutions presented in Appendix D.
For example, the first solution  in Table 9 (in the appendix)
results to the first two distinct cases of Table 2 and so on.

Note that in Table 2 we present only the vanishing vev's of each
particular solution. Substitution in the $F$-- and $D$--flatness
conditions, results to a number of constraints characterizing each
solution. These constraints are not presented in Table 2 but they
have been taken into account in the calculation of free
parameters. Specific examples will be presented later in Section
\ref{phen}. Due to the existence of free parameters, each solution
of Table 2  can in principle generate a number of
phenomenologically distinct cases.  We will see in a subsequent
section  how some of these free parameters are forced to obtain
zero vevs following the requirements of low energy phenomenology.

\underline{Case $({iv})$}\\ In a similar manner, we proceed also
in this case where $\bar\xi_1=\bar\xi_4 = 0$. Eqs. (\ref{ 9}),
(\ref{ 13}) lead to two sub--cases depending on whether $\phi_2$
is zero or not.\\
\begin{itemize}
\item $({\it iv})_a$
When $\phi_2=0$, the analysis proceeds in analogy with case
$(iii)$. Thus, we find eight  solutions at the tree--level which
cases 18-25 of Table 9.
\item $({\it iv_{b}})$
{}For the case  $\phi_2\not=0$, a tedious analysis leads to the
 unique tree--level solution
\ba
\xi_{2,3}=\bar\xi_i=\zeta_i=\bar\zeta_i=\bar F_{3R}=\bar F_{5R}=0
\ea
with $\Phi_2=4\,\Phi_{12}\,\bar\Phi_{12}\ne0$
\end{itemize}
This is also included as case 26 in the complete list of the
tree--level solutions of Table 9 in   Appendix D.   When
NR--contributions are taken into account various flat directions
are lifted and the total number of solutions is reduced to 4 which
are shown in Table 2 (cases (14)-(17)).

 Having obtained all consistent solutions, let us try to apply the
preliminary phenomenological discussion of the previous section.
We first point out that in eight of the  cases above,  all four
$\Phi_{12}$'s fields in the second row have zero vevs. Although
nothing can be definitely said until a complete analysis with
higher NR--terms is done, we consider them as less favored since
they leave all four sextet  fields massless at tree--level.
Another two solutions on the other hand, has all
$\ze_i=\bar\ze_i=0$. Again, according to our previous analysis, it
would be desirable to obtain a mass term for the second generation
at fourth order where a natural fermion mass hierarchy is
obtained. Such a solution should admit at least $\ze_1\ne 0$ and
$\langle
\bar{F}_{1R}\rangle = 0$, or ${\bar{\ze}_4}\not= 0$ and
$\langle\bar{F}_{2R}\rangle = 0$. {}From this point of view,  the
cases admitting non--zero vevs for some of the $\ze_i,\bar\ze_i$
are more preferable. {}Few of them leave only the fourplet
$\bar{F}_{1R}\not= 0$ to be interpreted as the second $SU(4)\times
SU(2)_R$ breaking higgs,
 (the other being definitely $F_{4R}$),
while there are several cases  with
$\langle\bar{F}_{3R}\rangle\not= 0$. Moreover, since in most of
the cases $\langle\bar{F}_{5R}\rangle = 0$, this latter field
together with $F_{4L}$, are suitable to accommodate  the third
generation fermions who may receive a tree--level mass term via
the Yukawa coupling $\bar{F}_{5R}F_{4L}{h}_{4}$.

\begin{table}
\newcounter{seci}\setcounter{seci}{0}
\newcommand{\nnu}[0]{\addtocounter{seci}{1}\arabic{seci}}
\begin{center}
\begin{tabular}{|r|c|c|c|c|c|c|}
\hline
&$\Phi_{12}s$&$\Phi_i$&$\xi_i,\bar\xi_i$&$\zeta_i,
\bar\zeta_i$&$\bar{F}_i$&f.p.
\\\hline
$\nnu$&$12,{12}^{-},\overline{12}^{-}$&$2,4,5$&$4,
{\bar 1},\{\bar 2\},{\bar 4}$&$3,
{\bar 3}$&$\{\bar 1\}, \{\bar 2\}, {\bar 5}$&$6$\\\hline
$\nnu$&$12,{12}^{-},\overline{12}^{-}$&$2,4,5$&$4,
{\bar 1},{\bar 4}$&$3,{\bar 3}$&$\{\bar 1\},
\{\bar 3\}, {\bar 5}$&$7$\\\hline
$\nnu$&$12,{12}^{-},\overline{12}^{-}$&$2,4,5$&
$4,{\bar 1},\{\bar 2\},{\bar 4}$&${\bar 1},
{\bar 3}$&$\{\bar 1\}, \{\bar 2\}, {\bar 5}$&$6$\\\hline
$\nnu$&$12,{12}^{-},\overline{12}^{-}$&$2,4,5$&$
4,{\bar 1},{\bar 4}$&${\bar 1},
{\bar 3}$&$\{\bar 1\}, \{\bar 3\}, {\bar 5}$&$7$\\\hline
$\nnu$&$12,\overline{12},{12}^{-},\overline{12}^{-}$&$2,4,5$&$4,
[\bar1],\{\bar 2\},{\bar 4}$&$3,
{\bar 3}$&$\{\bar 1\}, \{\bar 2\}, {\bar 5}$&$5$\\\hline
$\nnu$&$12,\overline{12},{12}^{-},\overline{12}^{-}$&$2,4,5$&$4,
[\bar 1],\{\bar 2\},{\bar 4}$&${\bar 1},{\bar 3}$&$\{\bar 1\},
\{\bar 2\}, {\bar 5}$&$6$\\\hline
$\nnu$&$12,{12}^{-},\overline{12}^{-}$& $2,5$&$4,{\bar 1},{\bar
4}$&$3, {\bar 1},{\bar 3}$&$\{\bar 1\}, \{\bar 3\} $&$7$\\\hline
$\nnu$&$12,{12}^{-},\overline{12}^{-}$&$2,5$&$4,{\bar 1}, \{\bar
2\}, {\bar 4}$&$1,3, {\bar 1},{\bar 3}$&$\{\bar 1\},\{\bar
2\},\{\bar 5\}$&$5$\\\hline
$\nnu$&$12,{12}^{-},\overline{12}^{-}$&$
2,5$&$4,{\bar 1},{\bar 4}$&$\{1\},3,
{\bar 1},{\bar 3}$&$\{\bar 2\}, \{\bar 3\},\{\bar 5\} $&$6$\\\hline
$\nnu$&$12,\overline{12},{12}^{-},\overline{12}^{-}$&$2,5$&$\{2\},4,
[\bar1],{\bar 4}$&$\{1\},3,{\bar 1},{\bar 3}$&$\{\bar 1\},
\{\bar 2\},\{\bar 5\} $&$4$\\\hline
$\nnu$&$12,\overline{12},{12}^{-},
\overline{12}^{-}$&$2,5$&$4,[\bar1],\{\bar 3\}, {\bar
4}$&$1,3,{\bar 1},{\bar 3}$&$\{\bar 1\}, \{\bar 2\},\{\bar 5\}
$&$4$\\\hline
$\nnu$&$12,{12}^{-},\overline{12}^{-}$&$2,4$&$4,{\bar 1},{\bar
4}$&$ 2,3,4,{\bar 2},{\bar 3},{\bar 4}$&$ \{\bar 1\},\{\bar
3\},{\bar 5}$&$5$\\\hline
$\nnu$&$12$&$2 $&$2,3,4,{\bar 1},\bar
2,{\bar 3},\bar 4$&$1,2,3,4, {\bar 1},{\bar 2},{\bar 3},{\bar
4}$&$ \{\bar 2\},\{\bar 3\},{\bar 5}$&$6$\\\hline
$\nnu$&$12,\overline{12},{12}^{-},\overline{12}^{-}$&
$2,4,5$&${\bar 1},\{\bar 2\},{\bar 4}$&$3,{\bar 3}$&$\{\bar 1\},
{\bar 5}$&$9$\\\hline
$\nnu$&$12,\overline{12},{12}^{-},\overline{12}^{-}$&
$2,4,5$&${\bar 1},\{\bar 2\},{\bar 4}$&${\bar 1},{\bar 3}$&$\{\bar
1\}, {\bar 5}$&$9$\\\hline
$\nnu$&$12,\overline{12},{12}^{-},\overline{12}^{-}$& $2,5$&${\bar
1},\{\bar 2\},{\bar 4}$&$3,{\bar 1},{\bar 3}$& $\{\bar 1\}$&$9$
\\\hline
 $\nnu$&$$&$$&$2,3,{\bar1},{\bar2},{\bar3},{\bar4}$&$1,2,3,4,
{\bar1},{\bar2},{\bar3},{\bar4}$&$\{\bar2\},{\bar3},{\bar5}$&$7$\\
\hline
\end{tabular}
\label{tab:allflat} \caption{{ The solutions to the $F$-- and
$D$--flatness equations with contributions of NR--terms up to
sixth order. The fields appearing in the table have zero vevs.
Those appearing in curly brackets $\{\}$ are forced to have zero
vevs form NR--contributions, while those in square brackets $[]$
are set to zero to ensure the existence of at least one massless
Higgs doublet. In the last column f.p. stands for the number of
free parameters. }}
\end{center}
\end{table}

\section{Higgs fields  and fermion mass textures \label{phen} }

We start our phenomenological analysis of the string model with
the discussion on the Higgs sector. Clearly, all the consistent
solutions of the flat directions considered in the previous
section automatically ensure the existence of one Higgs pair in
$({\bf4},{\bf1},{\bf2})+({\bf\bar{4}},{\bf1},{\bf2})$ to break the
$SU(4)\times SU(2)_R$ symmetry. Our next task is the securing  of
a massless pair of $SU(2)_L$ Higgs doublets in order to break the
electroweak symmetry. It suffices the existence of only one
massless Higgs bidoublet $(1,2,2)$, since after the first stage of
symmetry breaking two electroweak doublets with the correct
quantum numbers arise
\ba
h({\bf1},{\bf2},{\bf2})\ra h^u({\bf1},{\bf2},\frac 12) +
h^d({\bf1},{\bf2},-\frac 12) \label{dhd}
\ea
The Higgs matrix receives the following contributions from the
available tree--level superpotential couplings
\begin{equation}
m_{h} =  \left(\begin{array}{cccc}
  0   &   0   &  \xi_1  & \xi_4\\
  0   &   0   & \bar\xi_4 &\bar\xi_1\\
\xi_1 &\bar\xi_4&\bar\Phi_{12}&\phi_2\\
\xi_4 &\bar\xi_1&\phi_2&\Phi_{12}
\end{array}\right),
 \label{higgs}
\end{equation}
No further contributions to the Higgs matrix exist up to sixth
order. We will explore the eigenvalues of the above matrix in
conjunction with the flatness solutions discussed in the previous
section. In order to have at least one zero eigenvalue, the
determinant of $m_h$ should be zero
\ba
{\rm Det}(m_h)&=&
(\xi_1\bar\xi_1-\xi_4\bar\xi_4)^2 = 0 \label{detmh}
 \ea
 We notice that the determinant of the Higgs matrix does not depend
 on the fields vevs $\Phi_{12},\bar\Phi_{12}$ and $\phi_2$.

 We now come to the particular flat directions of Table 2. We observe
that  13 solutions arising from cases $(iii), (iv)$ have
automatically $\bar\xi_1=\bar\xi_4 = 0$. In the remaining 4
solutions the additional constraint $\bar\xi_1 = 0$ has to be
imposed  in order to ensure the existence of at least one massless
Higgs doublet. These are the cases (5,6,10,11) where the symbol
$[\bar 1]$ in the third column is used to declare the Higgs matrix
constraint on the singlet vev $\bar\xi_1$.

By inspection of the $D$--flatness equations (\ref{d1})-(\ref{d5})
we infer that two pairs of bidoublets are always massive. Going to
specific cases we find that the Higgs matrix   in solutions (1-13)
has exactly two zero eigenvalues corresponding to the pure states
$h_{2}, h_4$, while rest are massive. Solutions (14-16) have two
massless bidoublets. These are $h_2$ and the combination
$h'\propto-\xi_4 h_3 + \xi_1 h_4$
 The remaining solution (17) has only one massless bidoublet
 and more particularly $h_2$.

Each one of the above cases leads to a distinct phenomenological
model. It is convenient to classify them with respect to the
$({\bf\bar{4}},{\bf1},{\bf2})$ multiplet available for the Higgs
mechanism.
\begin{itemize}
\item There are seven cases, namely (1,3,5,6,8,10,11),
 where the only available field of this
type is $\bar{F}_{3R}$ since, as can be seen from the table,
$\langle\bar{F}_{(1,2,5)R}\rangle = 0$.
\item
In one single case (7) the $(\bar{4},1,2)$--Higgs in general can
be a linear combination of $\bar{F}_{2R},\bar{F}_{5R}$
\item
Only three solutions  admit
 $\langle\bar{F}_{1R}\rangle\not= 0$. These are (9,13,17).
\item
There are three solutions  where the higgs may be a linear
combination of $\bar{F}_{(2,3)R}$ (14,15) or $\bar{F}_{(2,3,5)R}$,
$SU(4)$ (16) multiplets.
\item
Finally, solutions (2,4,12)  admit only
 $\langle\bar{F}_{2R}\rangle\not= 0$.
\end{itemize}

Let us emphasize at this point that the above distinction between
the solutions together with the massless electroweak Higgs field
classification, is in accordance with common phenomenological
characteristics. {}For example, solutions of the first kind above
which impose $\langle\bar{F}_{3R}\rangle \not= 0$, have a rather
larger number of Yukawa couplings available for fermion mass
generation, making them more appealing. Also, from a further
inspection of the superpotential terms, the second class of
solutions with $\langle\bar{F}_{5R}\rangle \not= 0$ implies a mass
for the $h_4^u$ higgs via the tree--level term
$\langle\bar{F}_{5R}\rangle {F}_{4L}h_4$. This fact leaves only
one Yukawa coupling available for the up quarks, up to fifth order
making these solutions less interesting. In what follows, we will
work out in some detail some  representative cases from Table 2.\\

{\bf CASE 1}:\\ Let us start with the first solution in Table 2.
Along this flat direction, the following 15 fields are required to
have zero vevs \ba \Phi_{12}= \Phi_{12}^-= \bar\Phi_{12}^-=
\Phi_{2,4,5}= \xi_4= \bar{\xi}_{1,2,4} =\ze_{3}= \bar{\ze}_{3}=
\bar{F}_{1,2,5} = 0\ .
 \ea
 Among them, $\bar\xi_2$ and $\bar{F}_{5R}$
are constrained to have zero vevs from sixth order contributions.
Substituting the above condition to the full system of $D$-- and
$F$-- flatness equations we obtain a reduced system of 9 equations
for the remaining fields. These are \ba
 \bar\xi_2\,\bar\xi_3 + \ze_1^2 +\ze_2^2 + \ze_4^2 &=& 0\\
\bar\zeta_1^2 +\bar\zeta_2^2 +\bar\zeta_4^2 &=& 0\\
\xi_3\bar\xi_3\ + \zeta_1\bar\zeta_1+
\zeta_2\bar\zeta_2+\zeta_4\bar\zeta_4 &=&0\\
\zeta_2\bar\zeta_4+\zeta_4\bar\zeta_2 &=&0\label{coo}\\
 \frac{1}{2}|{\bar
F_{3R}}|^2 + |\xi_2|^2 + |\xi_3|^2 - |\bar\xi_3|^2 &=&0
\label{cooa}\\
    |{\bar F_{3R}}|^2  + 2\,|\xi_2|^2 + |\ze_1|^2 + |\ze_2|^2 +
     |\ze_4|^2- |\bar\ze_1|^2 -
        |\bar\ze_2|^2 - |\bar\ze_4|^2  &=& 0\label{coob}\\
    \frac{1}{2}|{\bar
F_{3R}}|^2 - |\xi_1|^2 &=& -\frac{\xi}{3}\label{cooc}\\
 |\xi_1|^2 +
|\bar\Phi_{12}|^2 &=& \frac{\xi}{2}\label{cood}\\
    |F_{4R}|^2 - |\bar F_{3R}|^2 &=&0 \label{cooe}\ea
 Taking into account
that the total number of fields available to obtain vevs are 30
(assuming that hidden sector fields do not develop vevs), we end
to a 6 parameter solution. This is the number of free parameters
(f.p.) presented in the last column of Table 2. As seen from the
above equations consistency of the solution requires a minimum
number of the remaining 15 fields
 \ba
\bar{F}_{3R},\; F_{4R},\; \Phi_{1,3},\;\xi_{1,2,3},\;
\bar\xi_{1,2,3},\; \ze_{1,2,4},\; \bar\ze_{1,2,4},\;\bar\Phi_{12}
\not=0\label{nzf}\ea to be non--zero. These are
$\xi_1\,,\bar\xi_3\,,{\bar F_{3R}}\,,{F_{4R}}$ and at least one of
$\ze_1,\ze_2,\ze_4$. The ${F_{4R}}$ and ${\bar F_{3R}}$ vevs are
not imposed by flatness but are required in order to obtain
$SU(4)\times SU(2)_R$ breaking.
 Thus the higgses in this case, in the notation of Section 2, are
  \ba
 F_{4R}\equiv H({\bf4},{\bf1},{\bf2}) ; &
\bar{F}_{3R} \equiv \bar{H}({\bf\bar 4},{\bf1},{\bf2})
 \label{Ha}
 \ea
 To explore the hierarchy of the fermion mass spectrum, we note first
 that since  $\langle \bar{F}_{3R}\rangle\not= 0$, the tree--level
Yukawa coupling $\bar{F}_{3R}F_{3L}h_2$ cannot be used  for a
fermion mass term.  Clearly, $F_{3L}$ is more appropriate for a
mirror partner for $\bar{F}_{5L}$.  Therefore ${\bar F_{5R}}$ and
$F_{4L}$ are suitable to accommodate a family and more
particularly the heaviest one as indicated by the tree--level
superpotential term $\bar{F}_{5R}F_{4L}h_4$. In this case we need
to impose $\langle\ze_2\rangle=0$, to avoid a mass term of the
form $\langle\ze_2\rangle F_{4L}\bar{F}_{5L}$.

Then, condition (\ref{coo}) results to two distinct cases, either
$\langle\ze_4\rangle = 0$, or $\langle\bar\ze_2\rangle= 0$, each
of them leading to a different phenomenological model. Although at
this level of calculation it cannot be decided which of the two
cases is appropriate, we consider the case $\langle\ze_4\rangle\ne
0$ as more favorable since it gives a tree--level mass term to a
pair of exotic states. Thus we choose to explore the case
$\langle\bar\ze_2\rangle= 0$.

To determine further the low energy parameters, we investigate the
$SU(4)$ breaking scale  constraints as well as the singlet vevs
entering the mass operators. {}From (\ref{cooc}),(\ref{cood}) it
follows that the $SU(4)$ breaking scale has a well defined upper
limit, determined exclusively from the $D$-term
\ba
|\bar{F}_{3R}|\le \sqrt{\frac{\xi}3} =
    \frac{g_{string}}{2\pi}M_{Pl}
\label{su4bs}
\ea
{}For perturbative values of $g_{string}$ (\ref{su4bs}) gives a
bound around the mass scale $10^{17}$GeV. Further, from
(\ref{coob}) we also conclude that $|\ze_1|< |\bar\ze_4|$. Up to
fifth order, we find the following Yukawa couplings suitable for
charged fermion masses
 \ba
\bar{F}_{5R}F_{4L} h_4 + \frac{\bar{\langle\ze_4
\rangle}}{M_{Pl}}\bar{F}_{2R}F_{2L}h_4 +
\frac{\langle\ze_1\rangle}{M_{Pl}}\bar{F}_{1R}F_{1L}h_4 \label{344}
 \ea
The last two terms appear at the fourth order, thus  an additional
mass parameter in their denominators appears. (In the Appendix B
the mass parameters in the denominators are omitted in order to
simplify the notation.) These two terms are obviously
hierarchically smaller than the first term which gives masses to
the third generation; further, taking into account the flatness
constraints, we infer that the second and first generations are
accommodated in ${\bar F}_{2R}, F_{2L}$ and ${\bar F}_{1R},F_{1L}$
respectively. From the constraints above, we are able to choose
$\ze_1 \ll \bar\ze_4$, so that we satisfy the mass hierarchies.
Moreover, this  implies that $\bar{F}_{3R}\sim
\bar\ze_4$. Recalling that $\bar{F}_{3R}$ plays the role of the
$SU(4)$--breaking higgs at the  scale $M_{GUT}$, we find that the
top--charm relation will determine further these vevs to be of the
order
\ba
 \frac{M_{GUT}}{M_{Pl}}\equiv
\frac{\langle\bar{F}_{3R}\rangle}{M_{Pl}}&\approx
 & \frac{m_c^0}{m_t^0}
\label{4S|CT}
\ea
It is worth noticing that this relation which correlates the
$SU(4)$ breaking scale with that of the  scale $M\sim M_{Pl}$
through the charm--top ratio at $M_{string}$, is in excellent
agreement with both, the flatness condition (\ref{su4bs}) as well
as the unification scale of the minimal unification scenario.  We
thus conclude that in the flat direction under consideration the
flavor assignments of the light standard model quarks and lepton
fields are as follows
 \ba
 F_{1L}:\; (u,d), (e, \nu_e)\; ; &
\bar{F}_{1R} :
                     \; u^c,d^c,e^c, \nu^c_e
\nonumber\\
 F_{2L}:\; (c,s), (\mu, \nu_{\mu})\; ; &
                      \bar{F}_{2R} :\; c^c,s^c,\mu^c,\nu^c_{\mu}
                         \label{assign}\\
F_{4L}:\; (t,b), (\tau, \nu_{\tau})\; ; &
                     \bar{F}_{5R} :\; t^c,b^c,\tau^c,\nu^c_{\tau}
\nonumber
\ea

Up to now, we have a rather successful picture of the fermion mass
spectrum which is also in  agreement with the string constraints.
The above accommodation of the fermion generations and Higgs
fields leaves no  arbitrariness as far as the extra vector--like
states are concerned: these are $F_{3L}$ and $\bar F_{5L}$. The
only mass term available at tree--level using fields of  the
observable sector, is proportional to the singlet vev
$\langle\ze_2\rangle$ however this is zero in the present case.
Nevertheless, we observe that there are terms involving hidden
fields which may  acquire non--zero vevs and give a heavy mass to
the mirror particles. For example, this can be obtained with  a
non--zero vev of the combination  $\langle Z_3\bar Z_4\rangle
\not= 0$ while they are constrained by the $D$--flatness to have
equal vevs $|\langle Z_3\rangle |= |\langle\bar Z_4\rangle |\,$\
\footnote{A similar mechanism has also been used in the
flipped--$SU(5)$ case \cite{art}}.

We turn now to the neutrino sector. The three terms in (\ref{344})
imply also Dirac masses for the corresponding neutrinos with
initial conditions at $M_{string}$ being the same as those for the
up--quarks. Therefore a see--saw mechanism is necessary to bring
them down to experimentally acceptable scales. An available term
exists already in the tree--level superpotential, which couples
the right handed neutrino $\nu_5^c\sim \bar{F}_{5R}^c$ with the
singlet field $\bar{\ze}_3$ via the vev $\langle F_{4R}\rangle $.
This leads to a see--saw mechanism of the type discussed in
Section 2. If we wish to find a final solution within the
observable sector field vevs, however, a complete account of the
neutrino mass problem needs the calculation of even higher
non--renormalizable terms. Restricting ourselves to contributions
of NR--terms up to fifth order, the see--saw mechanism, in
principle, can be effective for all neutrino species only when
additional hidden fields acquire vevs. In this case it is easily
checked that the following additional terms are generated \ba A
\nu_5^c\phi_2 + B \nu_2^c\phi_2 + C \nu_2^c\bar\ze_3 + D
\nu_1^c\bar\ze_2 \label{hs-s} \ea arising from the hidden sector
non--renormalizable contributions: \ba
A=\langle{F_{4R}Z_4\bar{Z}_5}\rangle,\;
B=\langle{F_{4R}Z_4\bar{Z}_2}\rangle,\;
C=\langle{F_{4R}Z_2\bar{Z}_4}\rangle,\;
D=\langle{F_{4R}Z_1\bar{Z}_4}\rangle. \ea The terms (\ref{hs-s})
complete the mechanism for all three flavours of neutrinos and
lead to an extended see--saw of the type (\ref{se-sa}). We note
however that the inclusion of the above hidden vevs requires a
re--examination of the flatness conditions.

We come now to the fields having fractional charges. Since no
 $\pm 1/2$-charge particle has been observed, the doublet states
 $X_{L,R}$ should also receive masses at some point, presumably
higher than the electroweak scale. If we write the doublet
$X_{iR}=(\chi_i^+,\chi_i^-)$, then a possible mass term would be
of the form:
\ba
W_{X} &=& \langle \phi \rangle
\epsilon_{ab}X_{iR}^aX_{jR}^b \nonumber\\
 &=&  \langle \phi \rangle (\chi_i^+\chi_j^- -\chi_i^-\chi_j^+)
\label{exmass}
\ea
where  $\epsilon_{ab}$ is the $SU(2)$ antisymmetric tensor. $\phi$
can be any combination of fields acquiring vevs resulting to an
effective singlet along the neutral direction. A similar term can
also exist for the left  doublets $X_{L}$. Terms mixing left with
right doublets are also possible, however, they lead to masses of
the order of the electroweak scale and are not of interest here.
At the tree--level, in the present flat direction we have the
following mass terms \ba W_X^{3}= \langle {{\zeta}_{4}}\rangle
{{X}_{2R}} {{X}_{5R}}+ \langle {{\xi}_{2}}\rangle
{{X}_{2R}}{{X}_{6R}}+ \langle {{\zeta}_{1}}\rangle  {{X}_{1R}}
{{X}_{6R}} \ea and similarly there are two terms for the left
doublet fields. Notice that we have now made used of the non--zero
vev $\langle\ze_4\rangle \not= 0$ which, according to the flatness
conditions (\ref{coo}) implies $\langle\bar{\ze}_2 \rangle  = 0$.
There are still three and four pairs of right and left doublets
respectively needed to take masses at some scale well above $m_W$.
All possible terms up to fifth order, have been collected in
Appendix B. By an inspection of the relevant (to this flat
direction) terms up to this order, it follows that few of the
octet fields $Z_i,\bar{Z}_i$ of the hidden sector with non--zero
vevs, are adequate to make  all of them massive. As has already
been pointed out, however, this will require a re--examination of
the flatness conditions \cite{fw}. All the same, the observable
singlet vevs may prove to be sufficient if higher order
contributions are calculated. A similar term may also appear for
the two $SU(4)$ fourplets $H_4=({\bf4},{\bf1},{\bf1})$,
$\bar{H}_4= ({\bf\bar{4}},{\bf1},{\bf1})$.

There is finally the rather important issue concerning the triplet
fields related to the stability of the proton. Recall first from
the detailed analysis in Section 2 that the triplets live only in
the sextets and the fourplet Higgs fields. There are two types of
terms here to render them massive. In the present case, there is
only one mass term available for the sextet fields at three level,
namely $\bar\Phi_{12}D_3D_4$, while the terms $F_{4R}^2D_3$ and
$\bar{\ze}_{1,4}^2F_{4R}^2D_4$ offer additional couplings with the
uneaten Higgs triplet of $F_{4R}$. Thus, up to fifth order, three
triplet pairs remain light. Higher order NR--terms will make them
massive. In particular, an inspection of the related seventh order
non--renormalizable superpotential mass terms shows that there are
plenty of available couplings rendering all but one pair massive
\ba W_D^{\le 7}&=& D_1D_2 \xi_1Z_3\bar{Z}_3Z_5\bar{Z}_5+
       D_1D_3 \left[ \ze_4\xi_1Z_2\bar{Z}_2(1+\Phi_1)+
                     \xi_1Z_2\bar{Z}_2Z_5\bar{Z}_4\right]
                     \nonumber\\
   &+& D_1D_4\left[\ze_1^2Z_5\bar{Z}_5\xi_1+
                    \bar{\ze}_4^2\xi_1Z_5\bar{Z}_5\right].
\ea It can be checked that the only coloured field which remained
uncoupled is the triplet $d_3^c$ of the Higgs field $\bar{H}\equiv
\bar{F}_{3R}$ leading to a pair of massless triplets. This has to
do with the fact that fields arising from the second $b_3$ and
being charged under peculiar $U(1)_4$ factor make only few
non--zero Yukawa couplings with other fields. For the same reason,
quarks and lepton fields do not also have dangerous couplings with
this triplet field up to this order. At higher orders, singlet
fields with non--zero $U(1)_4$ charge are expected to form
NR--term mass terms for $d_3^c$ and its partner so that proton
decay could be avoided.

{\bf CASE 7}\\ Here  we have the following zero vevs
 \ba
 \Phi_{12}^{-},\bar\Phi_{12}^{-},\Phi_{2,5},\xi_4,\bar\xi_{1,4},
 \ze_3,\bar\ze_{1,3},\bar F_{1R},\bar F_{3R} = 0
 \ea
Following the steps of the analysis of the previous case we find
that flatness reduces to 10 equations and thus the number of free
parameters is 7. The $SU(4)$--breaking Higgs fields are $F_{4R}$
and  a linear combination of the fourplets $\bar{F}_{2R}$ and
$\bar{F}_{5R}$. We now restrict to use the case
$\langle\bar{F}_{2R}\rangle = 0$. Thus, the higgses are
\ba
 F_{4R}\equiv H({\bf4},{\bf1},{\bf2}) ; \bar{F}_{5R} \equiv
  \bar{H}({\bf\bar 4},{\bf1},{\bf2})
                         \label{Hb}
\ea
Since $\langle\ze_2\rangle\not= 0$, the trilinear term
$\bar{F}_{5L}F_{4L}\ze_2$ makes massive the extra vector--like
states. On the other hand the terms
\ba
\bar{F}_{5R}F_{4L}h_4 +
\bar{F}_{5L}F_{4L}\ze_2\ra (\langle\nu_5^c\rangle
h_4^u+\langle\ze_2\rangle\bar\ell_5)\ell_4 \label{mhd}
\ea
mix a combination of the  Higgs doublet in $h_4$ with $\bar
\ell_5$ in $\bar F_{5L}$ leaving massless the
 combination
 \ba
 h^u = \cos\phi h_4^u
-\sin\phi\bar\ell_5,\;
\tan\phi=\frac{\langle\ze_2\rangle}{\langle\nu_5^c\rangle}
\label{lhc}
 \ea
 Once we have determined the electroweak Higgs eigenstates we are
 in a position to examine the available fermion mass terms. As
 previously, we will analyze Yukawa couplings up to fifth order.
 It is natural to accommodate the third generation in
the representations arising from the  sector $b_3$; due to the
existing fermion hierarchy the heavy fermions are expected to
obtain their mass through the only available tree--level term
\ba
\bar{F}_{3R}F_{3L}h_2&\ra &
     \langle h_2^u\rangle(tt^c + \nu_{\tau}\nu_{\tau}^c)
     +\langle h_2^d\rangle  (b b^c+\tau\tau^c)
     \nonumber
\ea
Then the lighter generations receive masses from
non--renormalizable terms,
\ba
\langle
h_4\bar\ze_4(1+\Phi_1)\rangle \bar{F}_{2R}F_{2L}+ \langle
h_4\rangle \langle(\ze_1(1+\Phi_3)+\bar{F}_{5R}F_{4R})\rangle
\bar{F}_{1R}F_{1L} \label{homt}
 \ea
where denominators of proper powers of $M_{string}$ in the various
NR--contributions are omitted.  Taking into account (\ref{lhc}),
the first term of (\ref{homt}) becomes
 \ba
\bar{F}_{2R}F_{2L}h_4\bar\ze_4&\ra &
\cos\phi\langle\bar\ze_4h^{u\prime}\rangle
(Q_2u_2^c+\nu_2^c\ell_2)+ \langle h^d_4\rangle
(Q_2d_2^c+\ell_2e_2^c)
 \ea
and similarly for the other terms. Additional contributions may
arise when higher order NR--terms are taken into account.

The triplet mass matrix in the present case, receives
contributions from terms involving the above non--zero vevs.
Assuming the sextet decompositions $D_i=D_i^3+\bar{D}_i^3$ the
triplet matrix takes the following form in the basis
$D_1,D_2,D_3,D_4, \bar{d}_{4}^c$,
\ba
\begin{array}{c|ccccc}
&D_1^3&D_2^3&D^3_3&D_4^3&\bar{d}_{4}^c(F_{4R})\\ \hline
\bar{D}_1^3& 0&x&x&x&0\\
\bar{D}_2^3&x&0&0&F_{4R}\bar\ze_i^2&F_{4R}\bar\ze_i^2\\
\bar{D}_3^3&x&0&0&\bar{\Phi}_{12}&F_{4R}    \\
\bar{D}_4^3&
x&0&\bar{\Phi}_{12}&0&0              \\
\parbox{0.8cm}{$\bar{d^c}_{5R}$\\
$({\bar F_{5R}})$}&\bar{F}_{5R}\bar\ze_i^2&\bar{F}_{5R}&0&0&0
\end{array}
\ea
where $\bar\ze_i$ stands for  the non--vanishing vevs
$\bar\ze_2,\bar\ze_4$. The symbol $x$ in the first row and column
of the above matrix represents possible contributions from
NR--terms involving fields from the hidden sector. These are
\ba
W_D^{\mathcal H}&=&
D_1D_2(\bar\ze_2Z_3\bar{Z}_3Z_5\bar{Z}_4+\xi_1Z_3\bar{Z}_3Z_5\bar{Z}_5)
+D_1D_3\ze_4\xi_1Z_2\bar{Z}_2 \nonumber\\ &+&
D_1D_4(\bar\ze_2\bar\ze_4^2Z_5\bar{Z}_4
+\bar\ze_4^2\xi_1Z_5\bar{Z}_5 +\bar\ze_4^2\xi_1Z_5\bar{Z}_4)
\nonumber
\ea
As can be seen from the mass--matrix,  observable sector
contributions up to seventh order make all but one pair of the
coloured triplets massive. If hidden fields also acquire vevs then
all triplets could  become  massive.

{\bf CASE 13}\\ We briefly comment now on another characteristic
case of Table 2, namely solution (13). This case is distinguished
by two remarkable properties which are worth mentioning:\\ {\it
(i)} First we observe that {\it all coloured sextets} become
massive at tree--level. This can be seen from the mass formula
(\ref{mcs}) and the fact that only one of the four singlets
involved in the tree--level mass matrix is required to have a zero
vev (namely $\langle\Phi_{12}\rangle = 0$).
\\ {\it (ii)}
Second, we point out that  the two lighter generations are not
pure states, since they appear to mixing appears already at
tree--level. To see this, we check first from Table 2 that the
solution requires $\langle\bar{F}_{2,3,5}\rangle = 0$, thus the
$SU(4)\times SU(2)_R$ Higgs fields are now $F_{4R}$ and
$\bar{F}_{1R}$. A possible mass term for the mirror states  may
appear now at a higher order (unless -- as previously -- hidden
fields obtain non--zero  vevs). The fermion mass terms are in this
case
\ba
\bar{F}_{3R}F_{3L}\langle h_2\rangle +
 \bar{F}_{5R}\left(F_{4L}\langle h_4 \rangle+
 F_{1L}h_2 \langle \bar{F}_{1R}F_{4R}h_4\right)\rangle
\ea
Clearly, the right--handed fields leaving in $\bar{F}_{5R}$ mix
with both $F_{1L}$ and $F_{4L}$. The flavor assignments are now
\ba
F_{1L}:\; (u',d'), (e', \nu_e')\; ; &
         \bar{F}_{2R} :\; u^{c'},d^{c'},e^{c'},\nu^c_{e'}
\nonumber\\
 F_{2L}:\; (c',s'), (\mu', \nu_{\mu}')\; ; &
                  \bar{F}_{2R} :\; c^{c'},s^{c'},\mu^{c'},
\nu^{c'}_{\mu}
                         \label{assign2}\\
F_{3L}:\; (t,b), (\tau, \nu_{\tau})\; ; &
                         \bar{F}_{3R} :\; t^c,b^c,\tau^c,\nu^c_{\tau}
\nonumber
\ea
where primes are used to denote that there is mixing in the two
lighter generations. We should point out here that the third
family remains essentially decoupled due to the peculiar
properties of the fourth $U(1)$. Only very high order NR--terms
are possible to mix this family with the lighter ones. This fact
of course predicts smaller mixing angles between the third family
with the rest of the fermion spectrum in consistency with the
phenomenological expectations.

\section{A brief discussion on the role of the hidden sector fields}

Up to now, we dealt with solutions of the flatness conditions
considering only non--zero vevs for observable fields. In the
phenomenological analysis of the previous section, however, we
have seen that couplings involving only observable--sector field
vevs are not adequate to make all exotic particles massive. There
are mainly two important issues to be further investigated before
this model is confronted with the low energy physics world. First,
higher order NR--terms have to be calculated in order to find all
possible contributions to the mass matrices of fermions, triplets
and other fields presented in the previous section. On the other
hand, hidden fields can also play a very important role on the
determination of the true vacuum of the model. Since they carry no
charge, they can also develop non--zero vevs and contribute to the
masses of the light fields through their Yukawa interactions. This
fact has been clear already in the three examples used in the
previous section. Neutrino masses, exotic states, and few of the
triplet fields become massive only when hidden fields are
included. At the same time, in both cases, the additional terms
contribute also to the flatness conditions, thus the new non--zero
vevs have to be carefully chosen so that they define a consistent
$F$-- and $D$--flat direction. {}For such an investigation, one
has to modify the flatness solutions, starting again from the tree
level cases which are included in Appendix D. A systematic
analysis of this general case is possible, however, this goes
beyond the scope of the present work \cite{fw}. {}For
completeness, the $D-$flatness conditions, in the presence of
non--zero hidden field vevs is given in Appendix C. Moreover, the
$F-$flatness conditions are written with the hidden fields
contributions up to fourth order in the same Appendix. Higher
order NR--corrections with hidden as well as observable field vevs
are easily extracted from the terms presented in Appendix B. In
the following, we give a brief account of the possible solutions
the hidden fields may give to some of the unanswered questions of
the present string construction.

Several constraints have to be carefully derived before some of
the  hidden  representations acquire vevs. An important constraint
arises from the demand of existence of massless electroweak Higgs
doublets. Although up to sixth order we have found no extra
contributions to the Higgs doublet matrix, such terms may well
exist in higher order, in particular when hidden fields are
allowed to obtain non--zero vevs. It should be noted that due to
the existence of high vevs associated with the $U(1)_A$ breaking
scale $M_A\sim 10^{-1} M_{Pl}$ securing the existence of massless
electroweak doublets is not an easy task for any superstring
model. To be more specific, assume a generic form of doublet mass
term $$g \left(\frac{\Phi}{M_{Pl}}\right)^n \Phi\,h\,h$$ with
$\Phi$ representing a typical singlet field obtaining a vev of the
order $\langle \Phi\rangle \sim M_A$. It is easy to see then that
even a $n=14$ order NR--term would in principle produce higgs
masses above the electroweak scale. Of course the existence of
superfluous doublet fields --as is the case of the model under
consideration-- provides the hope that even at this high  order of
calculation there exist flat directions that preserve at least one
pair of  doublets massless.

Another  severe constraint arises from the necessity to keep the
large $SU(4)\times SU(2)_R$ breaking Higgs field $F_{4R}$
massless. Up to sixth order, this can be ensured if the following
combinations of vevs are zero
 \ba
     \bar\ze_3,      &      \bar Z_3 Z_4,
 &   \phi_2 Z_4\bar Z_5,  \nonumber\\
\phi_2 \bar Z_2 Z_4, & \bar\ze_2 Z_1 \bar Z_4,
 &  \xi_1 Z_1 \bar Z_5,    \nonumber\\
\Phi_3 \Phi_3 \bar Z_3 Z_4, & \bar\ze_1 \ze_3
 \bar Z_1 Z_4,
& \bar\ze_1 \xi_1\bar Z_1 Z_5.
\ea
One might think that the above constraints demand most of the
hidden fields $Z_i$, $\bar{Z}_j$ to obtain zero vevs.  We would
like to point out however that it is possible to have a condition
of the form $\langle Z_i\bar{Z}_j\rangle = 0$, while at the same
time both fields may have non--zero vevs, $\langle Z_i\rangle
\not= 0$ and $\langle\bar{Z}_j\rangle \not= 0$. This happens
whenever the fields $Z_i$ and $\bar{Z}_j$ obtain their non--zero
vevs in orthogonal to each--other directions.

When some of the $SU(8)$ hidden fields acquire non--zero vevs, the
$SU(8)\times U(1)'$ symmetry is broken to a smaller group.
However, independently of the number of the hidden states which
develop non--zero vevs, there is always at least one unbroken
$U(1)''$ generator left, which is in general a linear combination
of the $U(1)'$ and one of the generators of $SU(8)$. On the other
hand, we note that the maximum number of $U(1)$ factors which may
remain unbroken in this model is two. Indeed, it can be checked
that the breaking of the $SU(4)\times SU(2)_R$ symmetry on one
hand and the consistency of the flatness conditions on the other
require at least the fields $F_{4R}, \bar{F}_{1R}$ and
$\bar{\Phi}_{12},\xi_1$ to develop non--zero vevs. These vevs
break three of the five Abelian factors.

The survival of $U(1)$ symmetries in lower energies would imply
the  stability of lightest observable and/or hidden fields being
charged under these symmetries. In all flat directions which were
previously analyzed, when the various singlet fields  obtain their
vevs, they break four out of five $U(1)$ factors. Thus, only the
aforementioned $U(1)''$ remains at low energies whilst, as a
consequence the lightest hidden state  will be stable. This fact
has important cosmological implications which we now briefly
discuss:

The last few years there is accumulating evidence from
astronomical observations that the universe is dominated by
invisible non--baryonic matter. According to a recent proposal
\cite{c1,c2} the dark matter of the universe --which is expected
to be ten times more that the luminous one-- might be composed
from non--thermal superheavy states produced in the early universe
provided that the following two conditions are met: $i)$ candidate
particles $Y$ should have a lifetime longer that the age of the
universe, $\tau_Y \ge 10^{10} y$, and $ii)$ they should not reach
local thermal equilibrium with the primordial plasma. To avoid
this constraint while having the correct number of $Y$ to form the
cold dark matter of the universe, it was suggested that these
particles are created through the interaction of the vacuum with
the gravitational field. Their mass is found to be around $m_Y\sim
10^{13}GeV${\footnote{For a similar discussion on the role of the
hidden matter fields in other string models see also \cite{c3}.}.

In the  string vacua found in the previous sections, a number of
the hidden states $Z_i,\bar{Z}_i$  in the present string
construction receive masses  at scales which are of the order of
the string mass. There are few of $Z_i,\bar{Z}_i$ states however,
which remain in the massless spectrum to lower scales. It is
possible that in certain string vacua the lightest hidden state
has a mass in the range $M_Y\sim 10^{13}$GeV as required in the
above scenario. As an example, we construct here the octet mass
matrix  for the solution 1 of Table 2. In the basis $Z_{1,\dots ,
5}$, the contributions up to sixth order involving only the
non--zero vev observable fields give the following texture
\ba
\left( \begin{array}{ccccc}
      \xi_1\bar\ze_1&0&0&0&0\\
      0&\xi_1\bar\ze_4&0&0&0\\
      0&0&0&F_{4R}\bar{F}_{3R}&0  \\
      0&0&0&\bar{\ze}_2&0  \\
      0&0&0&0&\xi_1
\end{array}
\right) \label{hidma}
 \ea
 In this case, four out of five hidden octet/octet--bar pairs receive
masses of the order of the $U(1)_A$ breaking mass scale $M_A\ge
M_{GUT}$. If hidden fields are also allowed to obtain vevs, then,
$Z_{4,5}$ are further mixed via the mass terms
$\langle\xi_1Z_1\bar{Z}_1\rangle \bar{Z}_4Z_5+
\langle\Phi_3Z_1\bar{Z}_1\rangle \bar{Z}_5Z_4$. There is only one
massless state (namely $Z_3$) up to this order. It is expected
that higher order terms will  provide a higher order
NR--contribution and make the remaining lightest  hidden pair
massive at the right scale, which is of course much lower that the
mass scale $M_A$ of the other $Z_i$-fields, as required by the
above cosmological scenario.

 \section{Conclusions}
In this paper, we have worked out an $SU(4)\times SU(2)_L\times
SU(2)_R$ model derived in the context of the four dimensional free
fermionic formulation of the heterotic superstring. Choosing a set
of nine vectors of boundary conditions on the world--sheet fermion
phases and appropriate GSO projection coefficients, we derived a
three--generation model supplemented by a mirror family and just
the necessary Higgs representations to break the symmetry down to
the standard model. In addition to the observable gauge symmetry,
the string model possesses also five $U(1)$'s as well as a hidden
$SU(8)$ gauge group.  The model predicts the existence of new
states beyond those of the minimal supersymmetric standard model
massless spectrum. These involve a large number of neutral singlet
fields, coloured $SU(4)$--sextets, $SU(8)$-octet hidden fields and
exotic states with fractional charges under the standard
hypercharge definition.

The superpotential of the model has been derived taken into
account string selection rules.
All fermion mass terms
have been worked out in detail up to fifth order and the fermion
and Higgs mass matrix textures have been assiduously analyzed. The
model is found to possess an anomalous $U(1)$-symmetry implying
the generation of a $D$--term which is canceled by vacuum
expectation values of singlet fields along $D$-- and $F$--flat
directions of the superpotential.

To work out the phenomenological implications, we have performed a
detailed analysis of all $D$-- and $F$-- flat directions including
contributions of non--renormalizable superpotential terms up to
sixth order. At tree--level, 26 solutions to the flatness
conditions were found and were classified with respect to the
fields which are demanded to have zero vevs in each particular
case. It was further shown that, when sixth order NR--terms are
included the solutions reduce to seventeen. Each solution is
characterized by a the number of free parameters which are
essentially the field vevs left undetermined by the particular
solution.  Particular attention has been paid in the determination
of those conditions necessary to ensure the existence in the
massless spectrum of the $SU(4)$ breaking higgses and at least two
Higgs electroweak doublets in order to break the GUT and SM gauge
symmetries respectively. These conditions have been imposed as
additional constraints on the consistent $D$-- and $F$--flat
directions and all phenomenologically acceptable string vacua have
been determined.

Three distinct flat directions, characterized by their
$SU(4)$-higgs properties are investigated in detail and the
predictions of the corresponding field theory models are
discussed. \\
 {\it a}).
  The first of these predicts that the $M_{GUT}/M_{string}$--ratio
is related to the up quark mass ratio of the second and third
generations. The choice of the GUT breaking Higgs representations
leaves a sufficient number of Yukawa couplings which produce
naturally a hierarchical fermion mass spectrum for all three
generations through tree--level and fourth order
non--renormalizable superpotential terms. Further, an analysis of
the superpotential NR--terms up to sixth order shows that all but
one of the colour triplet fields become massive.  It is worth
noting that there are no dangerous proton decay operators up to
this order of calculation since the massless triplet pair does not
couple to ordinary matter fields up to the sixth order. The
absence of Yukawa couplings between this triplet and ordinary
matter fields may be attributed to the properties of peculiar
$U(1)$-symmetry of the specific string basis--vector generating
this particular state. It is likely however that higher order
terms may provide a heavy mass to the remaining colour triplet
pair. \\ {\it b}). In a second case analyzed in this work, a
similar hierarchical fermion mass pattern is found, while all
triplets become massive if in addition hidden fields are allowed
to acquire non--zero vevs. On the other hand, in contrast to the
first model analyzed in Section 7, here the GUT scale has only an
upper bound determined by the $U(1)_A$ breaking scale. \\ {\it
c}). Finally, a third effective field theory model is analyzed
where all colour sextet fields become massive at the tree--level.
This model has fewer Yukawa couplings available for masses,
however, additional fermion mass terms may arise from higher order
non--renormalizable terms.

A novel feature of the effective field theory is the existence of
an additional $U(1)$-symmetry which survives down to low energies,
and it is possessed by exotic states and the hidden sector fields.
It is argued that if the lightest of these states receives mass at
some intermediate scale, may play a role in the dark matter of the
universe.

In the present paper our phenomenological explorations have been
restricted mainly with respect to the following two issues: First,
while there exist various ways to define the electric charge
operator of the model (due to the existence of surplus U(1)
factors), only the standard hypercharge embedding has been
considered in the phenomenological analysis. We believe that it is
worth exploring also  different types of embedding although one
has to face difficulties mainly with low initial $\sin^2\theta_W$
values. Second, the investigation of flat directions has been
limited in the cases where only `observable' fields are allowed to
obtain non--zero vevs. Certainly, the inclusion of the hidden
states in the analysis will lead to a large number of new mass
terms, the breaking of the hidden symmetry and  modifications of
the  flat directions found in this work.  Yet, such a possibility
has to be compared with analogous investigations of higher order
NR--terms will may or may not prove sufficient to obtain realist
low energy effective theory.

\newpage


\section{Appendix A: The spectrum}
\newcommand{\oh}{\frac{1}{2}}
\newcommand{\av}[4]{\left(#1,#2,#3,#4\right)}
\newcommand{\nav}[3]{\left({\bf#1},{\bf#2},{\bf#3}\right)}
We collect here the massless observable and hidden superfield
spectrum of the model. Fermionic string models contain always an
untwisted --usually called Neveu--Schwarz (NS)--sector where all
world--sheet fermions are antiperiodic. In this sector, the GSO
projections leave always in the massless spectrum the multiplet
which contains the graviton, the dilaton and the two--index
antisymmetric tensor. The NS--sector includes also the gauge
bosons and other Higgs and singlet fields. Twisted (R) sectors
provide the generations and other matter fields.

The states are classified in four separate tables according to
their transformation properties under the various parts of the
gauge symmetry. In the first column of each table we give the
symbol of the representation as this is used in the text. In the
last column we show the relevant sector of the string basis. In
all other columns we exhibit the gauge group properties of the
states.   Thus, Table 3 contains the observable fields, which have
non--trivial transformation properties under the PS--symmetry.
These are obtained from the sectors $b_{1,2,3,4,5}$ and
$S+b_4+b_5$. They include the three generations, the higgses and
other fields. Table 4 includes the PS singlets with their charges
under the four $U(1)$ symmetries. In Table 5 we present the hidden
$SU(8)$ fields with the corresponding charges under the five
$U(1)$s. Finally, in Table 6 we collect all exotic states with
fractional charges under the standard hypercharge assignment.

\begin{centering}
\begin{table}[!b]
\begin{tabular}{|l|c|c|c|c|c|l|}
\hline field&$SU(4)\times {SU(2)}_L\times
{SU(2)}_R$&${U(1)}_1$&${U(1)}_2$& ${U(1)}_3$&${U(1)}_4$&sector\\
\hline ${\overline{F}}_{5L}$ &$\nav{{\bar4}}{2}{1}$ &$0$ & $0$ &
$-\oh$ &$0$  &$b_5$\\ ${\overline{F}}_{5R}$ &$\nav{{\bar4}}{1}{2}$
&$0$ & $0$ & $+\oh$ &$0$  &\\ $F_{4L}$ &$\nav{4}{2}{1}$       &$0$
& $+\oh$ & $0$ &$0$  &$b_4$\\ $F_{4R}$ &$\nav{4}{1}{2}$       &$0$
& $-\oh$ & $0$ &$0$  &\\ ${\overline{F}}_{3R}$
&$\nav{{\bar4}}{1}{2}$ &$-\oh$ & $0$ & $0$ &$+\oh$  &$b_3$\\
$F_{3L}$              &$\nav{4}{2}{1}$ &$+\oh$ & $0$ & $0$ &$+\oh$
&\\ ${\overline{F}}_{2R}$ &$\nav{{\bar4}}{1}{2}$ &$0$ & $0$ &
$+\oh$ &$0$  &$b_2$\\ $F_{2L}$ &$\nav{4}{2}{1}$       &$0$ & $0$ &
$+\oh$ &$0$  &\\ ${\overline{F}}_{1R}$ &$\nav{{\bar4}}{1}{2}$ &$0$
& $+\oh$ & $0$ &$0$  &$b_1$\\ $F_{1L}$
&$\nav{4}{2}{1}$       &$0$ & $+\oh$ & $0$ &$0$  &\\ $D_1$
&$\nav{6}{1}{1}$ &$0$ & $-1$ & $0$ &$0$  &$S$\\ $D_2$
&$\nav{6}{1}{1}$ &$0$ & $0$ & $-1$ &$0$  &\\ $D_3$
&$\nav{6}{1}{1}$ &$0$ & $+1$ & $0$ &$0$  &\\ $D_4$
&$\nav{6}{1}{1}$ &$0$ & $0$ & $+1$ &$0$  &\\ $h_1$
&$\nav{1}{2}{2}$ &$0$ & $0$ & $0$ &$+1$  &\\ $h_2$
&$\nav{1}{2}{2}$ &$0$ & $0$ & $0$ &$-1$  &\\ $h_3$
&$\nav{1}{2}{2}$ &$0$ & $+\oh$ & $+\oh$ &$0$  &$S+b_4+b_5$\\ $h_4$
&$\nav{1}{2}{2}$ &$0$ & $-\oh$ & $-\oh$ &$0$  &\\ \hline
\end{tabular}
\caption{Observable sector spectrum of the $SU(4)\times SU(2)_L\times
 SU(2)_R$ model. }
\end{table}
\end{centering}
\begin{centering}
\begin{table}
\begin{tabular}{|l|c|c|c|c|c|l|}
\hline
field&${U(1)}_1$&${U(1)}_2$&${U(1)}_3$&${U(1)}_4$&sector\\
\hline
$\Phi_A\ ,A=1,\dots,5$&$0$ & $0$ & $0$ &$0$  &$S$\\
$\Phi_{12}$&$0$ & $+1$ & $+1$ &$0$  &\\
$\Phi_{12}^{-}$&$0$ & $+1$ & $-1$ &$0$  &\\
${\overline\Phi}_{12}$&$0$ & $-1$ & $-1$ &$0$  &\\
${\overline\Phi}_{12}^{-}$&$0$ & $-1$ & $+1$ &$0$  &\\
$\zeta_i\ ,i=1,\dots,4$&$0$ & $-\oh$ & $+\oh$ &$0$  &$S+b_4+b_5$\\
$\overline{\zeta_i}\ ,i=1,\dots,4$&$0$ & $+\oh$ & $-\oh$ &$0$  &\\
$\xi_1$&$0$ & $-\oh$ & $-\oh$ &$-1$  &\\
$\xi_2$&$-1$ & $-\oh$ & $+\oh$ &$0$  &\\
$\xi_3$&$-1$ & $+\oh$ & $-\oh$ &$0$  &\\
$\xi_4$&$0$ & $+\oh$ & $+\oh$ &$-1$  &\\
$\overline{\xi}_1$&$0$ & $+\oh$ & $+\oh$ &$+1$  &\\
$\overline{\xi}_2$&$+1$ & $+\oh$ & $-\oh$ &$0$  &\\
$\overline{\xi}_3$&$+1$ & $-\oh$ & $+\oh$ &$0$  &\\
$\overline{\xi}_4$&$0$ & $-\oh$ & $-\oh$ &$+1$  &\\
\hline
\end{tabular}
\caption{Non--Abelian singlet fields and their ${U(1)}^4$ quantum
numbers (all these fields have zero ${U(1)}'$- charge). }
\end{table}
\end{centering}
\begin{centering}
\begin{table}
\begin{tabular}{|l|c|c|c|c|c|c|l|}
\hline
field&${U(1)}_1$&${U(1)}_2$&${U(1)}_3$&${U(1)}_4$&${U(1)}'$&$SU(8)$&sector\\
\hline
$Z_1$     &$0$ & $0$ & $+\oh$ &$+\oh$ &$+\oh$& $\bf 8$ &$b_1+b_6 (+\zeta)$\\
$\bar Z_1$&$0$ & $0$ & $+\oh$ &$+\oh$ &$-\oh$& $\bf \bar8$ &\\
$Z_2$     &$0$ & $+\oh$ & $0$ &$+\oh$ &$+\oh$& $\bf 8$ &$b_2+b_6 (+\zeta)$\\
$\bar Z_2$&$0$ & $+\oh$ & $0$ &$+\oh$ &$-\oh$& $\bf \bar8$ &\\
$Z_3$     &$-\oh$&$+\oh$ &$+\oh$ &$0$ &$+\oh$& $\bf 8$ &$b_3+b_6 (+\zeta)$\\
$\bar Z_3$&$+\oh$&$+\oh$ &$+\oh$ &$0$ &$-\oh$& $\bf \bar8$ &\\
$Z_4$     &$0$ & $0$ & $-\oh$ &$-\oh$ &$+\oh$& $\bf 8$ &$b_4+b_6 (+\zeta)$\\
$Z_5$     &$0$ & $0$ & $+\oh$ &$+\oh$ &$+\oh$& $\bf 8$ &\\
$\bar Z_4$&$0$ & $-\oh$ & $0$ &$-\oh$ &$-\oh$& $\bf \bar8$ &$b_5+b_6 (+\zeta)$\\
$\bar Z_5$&$0$ & $+\oh$ & $0$ &$+\oh$ &$-\oh$& $\bf \bar8$ &\\
\hline
\end{tabular}
\caption{Hidden sector states and their ${U(1)}^4\times {U(1)}' \times SU(8)$ quantum
numbers.}
\end{table}
\end{centering}

\begin{centering}
\begin{table}
\begin{tabular}{|l|c|c|c|c|c|c|l|}
\hline field&$SU(4)\times {SU(2)}_L\times
{SU(2)}_R$&${U(1)}_1$&${U(1)}_2$&
${U(1)}_3$&${U(1)}_4$&${U(1)}'$&sector\\ \hline $X_{1L}$
&$\nav{1}{2}{1}$ &$-\oh$ & $0$ & $+\oh$ &$0$ &$-1$ &$b_1+\alpha$\\
$X_{2L}$ &$\nav{1}{2}{1}$ &$+\oh$ & $0$ & $+\oh$ &$0$ &$-1$ &\\
$X_{1R}$ &$\nav{1}{1}{2}$ &$-\oh$ & $0$ & $-\oh$ &$0$ &$-1$ &\\
$X_{2R}$ &$\nav{1}{1}{2}$ &$+\oh$ & $0$ & $-\oh$ &$0$ &$-1$ &\\
$X_{3L}$ &$\nav{1}{2}{1}$ &$-\oh$ & $0$ & $-\oh$ &$0$ &$+1$
&$b_4+\alpha$\\ $X_{4L}$ &$\nav{1}{2}{1}$ &$-\oh$ & $0$ & $+\oh$
&$0$ &$-1$ &\\ $X_{3R}$ &$\nav{1}{1}{2}$ &$+\oh$ & $0$ & $-\oh$
&$0$ &$-1$ &\\ $X_{4R}$ &$\nav{1}{1}{2}$ &$+\oh$ & $0$ & $+\oh$
&$0$ &$+1$ &\\ $X_{5L}$ &$\nav{1}{2}{1}$ &$-\oh$ & $-\oh$ & $0$
&$0$ &$+1$ &$b_1+b_4+b_5+\alpha$\\ $X_{6L}$ &$\nav{1}{2}{1}$
&$+\oh$ & $-\oh$ & $0$ &$0$ &$+1$ &\\ $X_{5R}$ &$\nav{1}{1}{2}$
&$-\oh$ & $+\oh$ & $0$ &$0$ &$+1$ &\\ $X_{6R}$ &$\nav{1}{1}{2}$
&$+\oh$ & $+\oh$ & $0$ &$0$ &$+1$ &\\ $X_{7L}$ &$\nav{1}{2}{1}$
&$-\oh$ & $-\oh$ & $0$ &$0$ &$+1$ &$b_1+b_2+b_4+\alpha$\\ $X_{8L}$
&$\nav{1}{2}{1}$ &$-\oh$ & $+\oh$ & $0$ &$0$ &$-1$ &\\ $X_{7R}$
&$\nav{1}{1}{2}$ &$+\oh$ & $-\oh$ & $0$ &$0$ &$-1$ &\\ $X_{8R}$
&$\nav{1}{1}{2}$ &$+\oh$ & $+\oh$ & $0$ &$0$ &$+1$ &\\ $X_{9L}$
&$\nav{1}{2}{1}$ &$0$ & $-\oh$ & $+\oh$ &$-\oh$ &$+1$
&$b_2+b_3+b_5+\alpha$\\ $X_{10L}$ &$\nav{1}{2}{1}$ &$0$ & $-\oh$ &
$+\oh$ &$+\oh$ &$-1$ &\\ $X_{9R}$ &$\nav{1}{1}{2}$ &$0$ & $+\oh$ &
$-\oh$ &$-\oh$ &$-1$ &\\ $X_{10R}$ &$\nav{1}{1}{2}$ &$0$ & $+\oh$
& $-\oh$ &$+\oh$ &$+1$ &\\ $H_4$ &$\nav{4}{1}{1}$ &$-\oh$ & $0$ &
$0$ &$0$ &$+1$  &$S+b_2+b_4+\alpha$\\ ${\overline{H}_4}$
&$\nav{\bar{4}}{1}{1}$ &$+\oh$ & $0$ & $0$ &$0$ &$-1$  &\\ \hline
\end{tabular}
\caption{Exotic fractionally charged states and their $SU(4)\times
{SU(2)}_L \times {SU(2)}_R \times {U(1)}^4 \times {U(1)}'$ quantum
 numbers.}
\end{table}
\end{centering}
\clearpage

\section{Appendix B:  Non--renormalizable contributions}
In the first part of this Appendix we give a brief description of
the techniques used to calculate the tree--level and higher order
NR--superpotential terms of the model. In the second part we give
a list of the  non--renormalizable superpotential terms
 involving  mass
terms (up to fith order) and  $F$--flatness conditions (up to
sixth order).

1)\,The calculation of non--renormalizable contributions to the
superpotential in the context of free--fermionic formulation is a
straightforward but rather tedious task. The rules for calculation
of NR terms have been presented in \cite{KLN} while explicit
calculation for various models have been presented in
\cite{RTA,LN,FAR}.
 In general, a superpotential term involving
the chiral superfields $\Phi_1,\Phi_2,\dots,\Phi_N$ is
proportional to the correlator $$\Phi_1\,\Phi_2\dots\Phi_N\sim
\left\langle V^f_{\Phi_1}\,V^f_{\Phi_2}V^b_{\Phi_3}\dots
V^b_{\Phi_N} \right\rangle$$
 where $V^f_{\Phi}$ stands for the fermionic part of the vertex operator
 corresponding to the field $\Phi$ and $V^b_{\Phi}$ for the bosonic part.
 The correlators can be calculated using conformal field theory
 techniques developed in \cite{FM,ISI,KLL}. An important subtlety  is that
 in order to guarantee conformal invariance the bosonic vertex operators
 $V^b_{\Phi_4},\dots, V^b_{\Phi_N}$ need to be pictured changed to the zeroth picture.

 A superpotential term vanishes if the corresponding correlator
vanishes otherwise it leads to an ${\cal O}(1)$ coupling. There
are two systematic sources of zeros in the superpotential. The
first is group invariance, the second are the internal symmetries
associated with the fermionized compactified coordinates. The
former is obvious while latter has been explored in \cite{RT}
where a set of selection rules has been derived. Since these
selection rules help significantly to the reduction of candidate
superpotential terms we summarize here the basic results.

The fermions $\chi^1,\chi^2,\dots,\chi^6$ corresponding to the
compactified coordinates
 can be bosonized as follows
\ba
 (\chi^1\pm \imath\chi^2)/\sqrt{2}
&=& \exp\{\pm \imath S_{12}\}\nonumber\\ (\chi^3\pm
\imath\chi^4)/\sqrt{2} &=& \exp\{\pm \imath S_{34}\}\nonumber\\
(\chi^5\pm \imath\chi^6)/\sqrt{2} &=& \exp\{\pm \imath S_{56}\}
\nonumber \ea $N=2$ world--sheet superconformal symmetry implies
the existence of an extra current, which is expressed in terms of
$S_{ij}$ as follows \ba J(q)= \imath \partial_q
(S_{12}+S_{34}+S_{56}) \ea and which is promoted to three $U(1)$'s
generated by $S_{12},S_{34},S_{56}$. The relevant part of the
vertex operators has the form
\ba V^f_{-\frac 12}&\propto&
 e^{(\alpha-\frac 12) S_{12}} e^{(\beta-\frac 12)
 S_{34}} e^{(\gamma-\frac 12) S_{56}}
\nonumber\\ V^b_{-1}&\propto& e^{\alpha S_{12}} e^{\beta S_{34}}
e^{\gamma S_{56}} \label{vo} \ea
 where $ -\frac 12,(-1)$ are the
ghost numbers for fermions and bosons respectively. Physical
states can now separated in two types NS (untwisted) and
R(twisted). Each type can be further divided in three categories
(the three orbifold planes in the orbifold language). In the
notation of Eq. (\ref{vo}) the three categories of NS--fields have
charges $(\alpha,\beta,\gamma)$=$\{(1,0,0),(0,1,0),(0,0,1)\}$
while for R--fields $(\alpha,\beta,\gamma)$=$\{(0,\frac 12,\frac
12),(\frac 12,0,\frac 12),(\frac 12,\frac 12,0)\}$ respectively.

Using the terminology explained above we can derive a set of
selection rules based on the conservation of the three $U(1)$
charges $S_{12},S_{34},S_{56}$. These selection rules are
presented in Table \ref{tnr} for NR--terms up to ninth order. The
notation we use is to write in square brackets the allowed
partition of fields in each of the categories for given order $N$.
The allowed field type (NS or R) appears as a subscript. As an
example let us explain the allowed fifth order couplings. From the
table we read $[3_R,2_R,0]$ when the number of NS--fields is zero
and $[2_R,2_R,1_{NS}]$ when the number of NS--fields is one. The
first selection rule means that in any non--vanishing coupling
between twisted fields the three of them have to belong to a
common plane while the other two should both reside in one of the
other planes. In the case that one untwisted field participates in
the coupling, the twisted ones should reside in the other two
planes and there should be exactly two of them in each one. As
seen from the table, all $5^{th}$ order couplings which contain
more that one NS field, vanish.

In order to see the effect of the above selection rules we present
in Table \ref{esr} the number of couplings that are eliminated
(for $N\le5$) from this source in the model under consideration.
We also present the number of couplings surviving group invariance
and the final number of non--vanishing superpotential couplings.
\begin{table}
\begin{center}
\begin{tabular}{|c|r|r|r|r|} \hline N& total
&\parbox{2.5cm}{selection rule \\invariants}&\parbox{2cm}{group
invariants}&final\\ \hline 3&73150&11719&372&66\\ \hline
4&595665&128928&339&34\\ \hline 5&16559487&2268256&10886&339\\
\hline
\end{tabular}
\caption{\label{esr}The total number of
candidate superpotential terms (for N=3,4,5) and their number
after application of the selection rules, group invariance and the
final number after complete evaluation of the correlators for the
model under consideration.}
\end{center}
\end{table}
Going further to the evaluation of correlators one finds another
source of zeros. These are the Ising type correlators arising due
to the existence of non--trivial left--right paired world--sheet
fermions. For tree--level couplings, the non--vanishing Ising
correlators are
 \ba \langle\sigma_{+}\sigma_{+}\rangle\;\; ,
  \langle\sigma_{-}\sigma_{-}\rangle\;\; ,
\langle\sigma_+\sigma_-f\rangle\;\; , \langle\sigma_+\sigma_-\bar
f\rangle \ea for higher order terms one can follow the rules of
\cite{ISI}.

The whole problem of deriving the superpotential terms can be
automated using a computer program \cite{RTHESIS}. The selection
rules are initially used to reduce the number of candidate
couplings, then group invariance is checked and finally all Ising
type correlators are evaluated. The whole calculation takes a few
seconds on a personal computer for $N=5$ and comparable time for
selected $N=6$ couplings.

Using this program we have calculated non--vanishing
superpotential couplings.

\begin{sidewaystable}
\begin{tabular}{|l|c|c|c|c|c|}
\hline N&0 {\small NS}& 1 NS& 2 NS & 3 NS & 4 NS \\ \hline
3&$[1_R, 1_R, 1_R]$&$[2_R+1_{NS},0,0]$& - &$[1_{NS}, 1_{NS},
1_{NS}]$&$\cdot$
\\\hline
4&$[2_R,2_R,0]$& - & -& -&-
\\\hline
5&$[3_R,1_R,1_R]$&$[2_R,2_R,1_{NS}]$& -&-&-
\\\hline
6&$[4_R,2_R,0]$&$[3_R,1_R+1_{NS},1_R]$&$[2_R,2_R,2_{NS}]$&$\cdot$&$
\cdot$
\\
&$[2_R,2_R,2_R]$&&&&\\\hline
7&$[5_R,1_R,1_R]$&$[4_R,2_R,1_{NS}]$&$[3_R,1_R+2_{NS},1_R]$&$[2_R,
2_R,3_{NS}]$&-

\\
 &$[3_R,3_R,1_R]$&$[4_R,2_R+1_{NS},0]$&$[3_R,1_R+1_{NS},
1_R+1_{NS}]$& &
\\
&  &$[2_R,2_R,2_R+1_{NS}]$& & &\\\hline
8 & $[6_R,2_R,0]$&$[5_R,1_R+1_{NS},1_R]$&$[4_R,2_R,2_{NS}]
$&$[3_R,1_R,1_R+3_{NS}]$&$[2_R,2_R,4_{NS}]$
\\
 &$[4_R,4_R,0]$&$[3_R,3_R,1_R+1_{NS}]$&$[4_R,2_R+1_{NS},
 1_{NS}]$&$[3_R,1_R+1_{NS},1_R+2_{NS}]$&\\
 &$[4_R,2_R,2_R]$&$[3_R,3_R+1_{NS},1_R]$&$[4_R,2_R+2_{NS},0]$&&\\
&&&$[2_R+2_{NS},2_R,2_R]$&&\\
&&&$[2_R+1_{NS},2_R+1_{NS},2_R]$&&\\\hline
\end{tabular}
\caption{\label{tnr}Non--vanishing superpotential couplings up to
8$^{th}$ order.}
\end{sidewaystable}

2)Here we present fifth and sixth order NR--contributions to the
superpotential. For finiteness we list only terms related to
fermion masses and flatness conditions.

 {\bf a})The 5th order
superpotential  terms involving masses for observable fields are:
\\
\begin{eqnarray}
w_5&=&
\bar{F}_{2R}F_{2L}h_4\bar\ze_4\Phi_1+\bar{F}_{1R}F_{1L}\ze_1h_4\Phi_3+
      \bar{F}_{5R}F_{4R}\bar{F}_{1R}F_{1L}h_4\nonumber\\
   &+&
 \bar{F}_{5L}^2\left( ( F_{4L}^2+F_{1L}^2+
\bar{F}_{1R}^2)\bar\Phi_{12}^-+F_{4R}^2\Phi_{12}
+ D_1 (h_3^2+\bar\xi_1\xi_4)+D_3(\ze_i^2+\xi_2\bar\xi_3)\right)\nonumber\\
   &+&
  \bar{F}_{5L}\left(\bar{F}_{5R}\left( D_1\bar\ze_2h_3+D_3\ze_2h_4\right)
       +  F_{4L}\bar{F}_{2R}F_{2L}h_4 + F_{4R}
  (\bar{F}_{2R}F_{2L}\bar\ze_3 + \bar{F}_{1R}F_{1L}\ze_2)\right)\nonumber\\
   &+&
\bar{F}_{5R}^2\left( ( F_{4L}^2+F_{1L}^2+
\bar{F}_{1R}^2)\bar\Phi_{12}+F_{4R}^2\Phi_{12}^-
+ D_1 (\bar\ze_i^2+\bar\xi_2\xi_3)+
D_3(h_4^2+\xi_1\bar\xi_4)\right)\nonumber\\
   &+&
         \bar{F}_{5R} F_{4L}\bar{F}_{3R}F_{3L}\xi_1
       + F_{4L}F_{4R} \left( D_2\ze_3h_3 + D_4\bar\ze_3h_4\right)
\nonumber\\
   &+&
 F_{4L}^2\left( (\bar{F}_{2R}^2+F_{2L}^2)
\bar\Phi_{12} + D_2(\ze_i^2+\xi_2\bar\xi_3)
       + D_4 (h_4^2 +\xi_1\bar\xi_4)\right)\nonumber\\
   &+&
           F_{2R}^2\left( (\bar{F}_{1R}^2+F_{1L}^2)\bar\Phi_{12} +
 D_1(\bar\ze_i^2+\bar\xi_2\xi_3)+
D_3(h_4^2+\xi_1\bar\xi_4)\right)\nonumber\\
   &+&
       F_{2L}^2 \left((\bar{F}_{1R}^2 + F_{1L}^2)\bar\Phi_{12}+
D_1(\bar\ze_i^2+\bar\xi_2\xi_3)+D_3(h_4^2+\xi_1\bar\xi_4)\right)\nonumber\\
   &+&
F_{4R}^2\left( (\bar{F}_{2R}^2 + F_{2L}^2)\Phi_{12}^-
+D_4(\bar\ze_i^2 +\bar\xi_2\xi_3)
           +D_2 (h_3^2+\bar\xi_1\xi_4)\right)\nonumber\\
   &+&
      \left( F_{1L}^2+\bar{F}_{1R}^2\right)
  \left( D_2(\ze_i^2+\xi_2\bar\xi_3)+D_4(h_4^2+\xi_1\bar\xi_4)\right)
\end{eqnarray}

 {\bf b})The 5th order
superpotential  terms involving masses for exotic  fields are:
\begin{eqnarray}
w_5'&=& {\overline{F}_{5R}} {{F}_{4R}} {{\Phi}_{4}} {{X}_{1R}} {{X}_{6R}}+
        {\overline{F}_{5R}} {{F}_{4R}} {{\Phi}_{5}} {{X}_{1R}} {{X}_{6R}}+
   {\overline{F}_{5R}} {{F}_{4R}} {{\zeta}_{3}} {{X}_{9R}} {{X}_{10R}}+
{\overline{\xi}_{2}} {\overline{Z}_{1}} {{Z}_{4}} {{X}_{4L}} {{X}_{5L}}
 \nonumber\\&+ &
   {{F}_{4R}} {\overline{F}_{2R}} {{\Phi}_{4}} {{X}_{1R}} {{X}_{8R}}+
   {{F}_{4R}} {\overline{F}_{2R}} {{\Phi}_{5}} {{X}_{1R}} {{X}_{8R}}+
{{F}_{4R}}{\overline{F}_{2R}}{\overline{\zeta}_{3}} {{X}_{6L}} {{X}_{8L}}+
{\overline{\xi}_{2}} {{Z}_{2}} {\overline{Z}_{4}} {{X}_{1L}} {{X}_{7L}}
\nonumber\\&+ &
   {{F}_{4R}} {\overline{F}_{2R}}
 {\overline{\zeta}_{3}} {{X}_{5R}} {{X}_{7R}}+
{{F}_{4R}} {\overline{F}_{1R}} {{\zeta}_{4}} {{X}_{3R}} {{X}_{5R}}+
{{F}_{4R}} {\overline{F}_{1R}} {\overline{\zeta}_{4}} {{X}_{4L}} {{X}_{6L}}+
  {{F}_{4R}} {\overline{F}_{1R}} {{\xi}_{2}} {{X}_{3R}} {{X}_{6R}}
  \nonumber\\&+ &
{{F}_{4R}} {\overline{F}_{1R}} {\overline{\xi}_{2}} {{X}_{4L}} {{X}_{5L}}+
 {{\Phi}_{1}} {{\zeta}_{2}} {\overline{\xi}_{2}} {{X}_{7L}} {{X}_{8L}}+
   {{\Phi}_{1}} {\overline{\zeta}_{2}} {{\xi}_{2}} {{X}_{7R}} {{X}_{8R}}+
{{\Phi}_{1}} {\overline{\zeta}_{2}} {\overline{\xi}_{3}} {{X}_{7L}} {{X}_{8L}}
\nonumber\\&+ &
{{\Phi}_{3}} {\overline{\zeta}_{3}} {{\xi}_{2}} {{X}_{3R}} {{X}_{4R}} +
{{\Phi}_{3}} {{\zeta}_{3}} {{\xi}_{3}} {{X}_{3R}} {{X}_{4R}}+
{{\Phi}_{3}} {{\zeta}_{3}} {\overline{\xi}_{2}} {{X}_{3L}} {{X}_{4L}}+
{{\Phi}_{3}} {\overline{\zeta}_{3}} {\overline{\xi}_{3}} {{X}_{3L}} {{X}_{4L}}
\nonumber\\&+ &
{{\Phi}_{4}} {{Z}_{1}} {\overline{Z}_{4}} {{X}_{3R}} {{X}_{5R}}+
{{\Phi}_{4}} {\overline{Z}_{2}} {{Z}_{4}} {{X}_{2L}} {{X}_{7L}}+
{{\Phi}_{4}} {{Z}_{4}} {\overline{Z}_{5}} {{X}_{2L}} {{X}_{5L}}+
{{\Phi}_{4}} {{Z}_{5}} {\overline{Z}_{4}} {{X}_{2R}} {{X}_{5R}}
\nonumber\\&+ &
 {{\Phi}_{5}} {{Z}_{1}} {\overline{Z}_{4}} {{X}_{3R}} {{X}_{5R}}+
{{\Phi}_{5}} {\overline{Z}_{2}} {{Z}_{4}} {{X}_{2L}} {{X}_{7L}}+
{{\Phi}_{5}} {{Z}_{4}} {\overline{Z}_{5}} {{X}_{2L}} {{X}_{5L}}+
 {{\Phi}_{5}} {{Z}_{5}} {\overline{Z}_{4}} {{X}_{2R}} {{X}_{5R}}
\nonumber\\&+ &
 {{\zeta}_{1}} {{Z}_{2}} {\overline{Z}_{4}} {{X}_{1R}} {{X}_{8R}}+
{\overline{\zeta}_{1}} {{Z}_{2}} {\overline{Z}_{4}} {{X}_{2L}} {{X}_{7L}}+
{{\zeta}_{2}} {{Z}_{5}} {\overline{Z}_{4}} {{X}_{9R}} {{X}_{10R}}+
{\overline{\zeta}_{2}} {{Z}_{1}} {\overline{Z}_{4}} {{X}_{3L}} {{X}_{2L}}
\nonumber\\&+ &
{\overline{\zeta}_{2}} {{Z}_{1}} {\overline{Z}_{4}} {{X}_{4R}} {{X}_{1R}}+
 {{\zeta}_{3}} {\overline{Z}_{2}} {{Z}_{4}} {{X}_{6L}} {{X}_{8L}}+
 {{\zeta}_{3}} {\overline{Z}_{2}} {{Z}_{4}} {{X}_{5R}} {{X}_{7R}}+
{\overline{\zeta}_{3}} {{Z}_{4}} {\overline{Z}_{5}} {{X}_{9L}} {{X}_{10L}}
\nonumber\\&+ &
 {{\zeta}_{4}} {\overline{Z}_{1}} {{Z}_{4}} {{X}_{3R}} {{X}_{5R}}+
{\overline{\zeta}_{4}} {\overline{Z}_{1}} {{Z}_{4}} {{X}_{4L}} {{X}_{6L}}+
 {{\xi}_{2}} {\overline{Z}_{1}} {{Z}_{4}} {{X}_{3R}} {{X}_{6R}}+
{{\xi}_{2}} {{Z}_{2}} {\overline{Z}_{4}} {{X}_{2R}} {{X}_{8R}}
\nonumber\\&+ &
{{\xi}_{1}} {{Z}_{1}} {\overline{Z}_{5}} {{X}_{3L}} {{X}_{2L}}+
{{\xi}_{1}} {{Z}_{1}} {\overline{Z}_{5}} {{X}_{4R}} {{X}_{1R}}+
 {{\xi}_{1}} {\overline{Z}_{2}} {{Z}_{5}} {{X}_{6L}} {{X}_{8L}}+
{{\xi}_{1}} {\overline{Z}_{2}} {{Z}_{5}} {{X}_{5R}} {{X}_{7R}}
\end{eqnarray}

{\bf c}) 5th and 6th order contributions to the $F-$flatness are
\footnote{For convenience, we include here some of the terms
listed above since they can contribute in both categories
depending on the generation assignment. }:
\begin{eqnarray}
w_5''&=&
   {{\Phi}_{1}} {{\zeta}_{4}} {{\xi}_{1}} {\overline{Z}_{2}} {{Z}_{2}}+
   {{\Phi}_{2}} {\overline{F}_{5R}} {{F}_{4R}} {{Z}_{4}} {\overline{Z}_{5}}+
   {{\Phi}_{2}} {{F}_{4R}} {\overline{F}_{2R}} {\overline{Z}_{2}} {{Z}_{4}}+
   {{\Phi}_{3}} {\overline{\zeta}_{1}} {{\xi}_{1}} {\overline{Z}_{1}} {{Z}_{1}}
\nonumber\\ & &\mbox{}+
   {\overline{\Phi}_{12}} {\overline{F}_{5R}} {\overline{F}_{5R}}
   {\overline{F}_{1R}} {\overline{F}_{1R}}+
   {\overline{\Phi}_{12}} {\overline{F}_{2R}} {\overline{F}_{2R}}
   {\overline{F}_{1R}} {\overline{F}_{1R}}+
   {\overline{\Phi}_{12}^{-}} {\overline{Z}_{2}} {\overline{Z}_{2}}
   {{Z}_{4}} {{Z}_{4}}+
   {\overline{\Phi}_{12}^{-}} {{Z}_{4}} {{Z}_{4}} {\overline{Z}_{5}}
   {\overline{Z}_{5}}
\nonumber\\ & &\mbox{}+
   {{\Phi}_{12}^{-}} {\overline{F}_{5R}} {\overline{F}_{5R}} {{F}_{4R}}
   {{F}_{4R}}+
   {{\Phi}_{12}^{-}} {{F}_{4R}} {{F}_{4R}} {\overline{F}_{2R}}
   {\overline{F}_{2R}}+
   {{\Phi}_{12}^{-}} {{Z}_{1}} {{Z}_{1}} {\overline{Z}_{4}}
   {\overline{Z}_{4}}+
   {{\Phi}_{12}^{-}} {{Z}_{5}} {{Z}_{5}} {\overline{Z}_{4}} {\overline{Z}_{4}}
\nonumber\\ & &\mbox{}+
   {{F}_{4R}} {\overline{F}_{2R}} {\overline{\zeta}_{3}}
   {{Z}_{2}} {\overline{Z}_{4}}+
   {{F}_{4R}} {\overline{F}_{1R}} {\overline{\zeta}_{2}}
   {{Z}_{1}} {\overline{Z}_{4}}+
   {{F}_{4R}} {\overline{F}_{1R}} {{\xi}_{1}} {{Z}_{1}}
   {\overline{Z}_{5}}+
   {\overline{\zeta}_{2}} {\overline{Z}_{1}} {{Z}_{1}}
   {{Z}_{4}} {\overline{Z}_{4}}
\nonumber\\ & &\mbox{}+
   {{\zeta}_{3}} {\overline{Z}_{2}} {{Z}_{2}} {{Z}_{4}}
   {\overline{Z}_{4}}+
   {{\xi}_{1}} {\overline{Z}_{1}} {{Z}_{1}} {{Z}_{4}}
   {\overline{Z}_{5}}+
   {{\xi}_{1}} {\overline{Z}_{2}} {{Z}_{2}} {{Z}_{5}} {\overline{Z}_{4}}
\end{eqnarray}
\begin{eqnarray}
w_6&=&
{{\Phi}_{1}} {{\Phi}_{1}} {{\zeta}_{4}} {{\xi}_{1}} {\overline{Z}_{2}} {{Z}_{2}}+
{{\Phi}_{1}} {{F}_{4R}} {\overline{F}_{2R}} {\overline{\zeta}_{3}} {{Z}_{2}}{\overline{Z}_{4}}+
{{\Phi}_{1}} {{\zeta}_{3}} {\overline{Z}_{2}} {{Z}_{2}} {{Z}_{4}} {\overline{Z}_{4}}+
{{\Phi}_{1}} {{\xi}_{1}} {\overline{Z}_{2}} {{Z}_{2}} {{Z}_{5}} {\overline{Z}_{4}}
\nonumber\\ & &\mbox{}+
{{\Phi}_{2}} {{\Phi}_{5}} {{F}_{4R}} {\overline{F}_{2R}} {\overline{Z}_{2}} {{Z}_{4}}+
{{\Phi}_{3}} {{\Phi}_{3}} {{F}_{4R}} {\overline{F}_{3R}} {\overline{Z}_{3}} {{Z}_{4}}+
{{\Phi}_{3}} {{\Phi}_{3}} {\overline{\zeta}_{1}} {{\xi}_{1}} {\overline{Z}_{1}} {{Z}_{1}}+
{{\Phi}_{3}} {{F}_{4R}} {\overline{F}_{1R}} {\overline{\zeta}_{2}} {{Z}_{1}}
{\overline{Z}_{4}}
\nonumber\\ & &\mbox{}+
{{\Phi}_{3}} {{F}_{4R}} {\overline{F}_{1R}} {{\xi}_{1}} {{Z}_{1}} {\overline{Z}_{5}}+
{{\Phi}_{3}} {\overline{\zeta}_{2}} {\overline{Z}_{1}} {{Z}_{1}} {{Z}_{4}} {\overline{Z}_{4}}+
{{\Phi}_{3}} {{\xi}_{1}} {\overline{Z}_{1}} {{Z}_{1}} {{Z}_{4}} {\overline{Z}_{5}}+
{{\Phi}_{4}} {\overline{\Phi}_{12}} {\overline{F}_{5R}}
{\overline{F}_{5R}} {\overline{F}_{1R}} {\overline{F}_{1R}}
\nonumber\\ & &\mbox{}+
{{\Phi}_{4}} {\overline{\Phi}_{12}} {\overline{F}_{2R}} {\overline{F}_{2R}}
{\overline{F}_{1R}} {\overline{F}_{1R}}+
{{\Phi}_{4}} {{\Phi}_{12}^{-}} {{Z}_{1}} {{Z}_{1}} {\overline{Z}_{4}} {\overline{Z}_{4}}+
{{\Phi}_{4}} {{F}_{4R}} {\overline{F}_{1R}} {\overline{\zeta}_{2}} {{Z}_{1}}{\overline{Z}_{4}}+
{{\Phi}_{4}} {{F}_{4R}} {\overline{F}_{1R}} {{\xi}_{1}} {{Z}_{1}} {\overline{Z}_{5}}
\nonumber\\ & &\mbox{}+
{{\Phi}_{4}} {\overline{\zeta}_{2}} {\overline{Z}_{1}} {{Z}_{1}} {{Z}_{4}} {\overline{Z}_{4}}+
{{\Phi}_{4}} {{\xi}_{1}} {\overline{Z}_{1}} {{Z}_{1}} {{Z}_{4}} {\overline{Z}_{5}}+
 {{\Phi}_{5}} {\overline{\Phi}_{12}} {\overline{F}_{2R}} {\overline{F}_{2R}} \overline{F}_{1R}}
 {\overline{F}_{1R}}+ {{\Phi}_{5}} {\overline{\Phi}_{12}^{-}}
 {\overline{Z}_{2}} {\overline{Z}_{2}{{Z}_{4}} {{Z}_{4}}
\nonumber\\ & &\mbox{}+
{{\Phi}_{5}} {\overline{\Phi}_{12}^{-}} {\overline{Z}_{2}} {\overline{Z}_{2}}{{Z}_{4}} {{Z}_{4}}+
{{\Phi}_{5}} {{\Phi}_{12}^{-}} {{F}_{4R}} {{F}_{4R}} {\overline{F}_{2R}} {\overline{F}_{2R}}+
{{\Phi}_{5}} {{F}_{4R}} {\overline{F}_{2R}} {\overline{\zeta}_{3}} {{Z}_{2}}{\overline{Z}_{4}}+
{{\Phi}_{5}} {{\zeta}_{3}} {\overline{Z}_{2}} {{Z}_{2}} {{Z}_{4}} {\overline{Z}_{4}}
\nonumber\\ & &\mbox{}+
{{\Phi}_{5}} {\overline{\Phi}_{12}^{-}} {\overline{Z}_{2}} {\overline{Z}_{2}}{{Z}_{4}} {{Z}_{4}}+
{{\Phi}_{5}} {{\Phi}_{12}^{-}} {{F}_{4R}} {{F}_{4R}} {\overline{F}_{2R}} {\overline{F}_{2R}}+
{{\Phi}_{5}} {{F}_{4R}} {\overline{F}_{2R}} {\overline{\zeta}_{3}} {{Z}_{2}}{\overline{Z}_{4}}+
{{\Phi}_{5}} {{\zeta}_{3}} {\overline{Z}_{2}} {{Z}_{2}} {{Z}_{4}} {\overline{Z}_{4}}
\nonumber\\ & &\mbox{}+
{{\Phi}_{5}} {\overline{\Phi}_{12}^{-}} {\overline{Z}_{2}} {\overline{Z}_{2}}{{Z}_{4}} {{Z}_{4}}+
{{\Phi}_{5}} {{\Phi}_{12}^{-}} {{F}_{4R}} {{F}_{4R}} {\overline{F}_{2R}} {\overline{F}_{2R}}+
{{\Phi}_{5}} {{F}_{4R}} {\overline{F}_{2R}} {\overline{\zeta}_{3}} {{Z}_{2}}{\overline{Z}_{4}}+
{{\Phi}_{5}} {{\zeta}_{3}} {\overline{Z}_{2}} {{Z}_{2}} {{Z}_{4}} {\overline{Z}_{4}}
\nonumber\\ & &\mbox{}+
{{\Phi}_{5}} {\overline{\Phi}_{12}^{-}} {\overline{Z}_{2}} {\overline{Z}_{2}}{{Z}_{4}} {{Z}_{4}}+
{{\Phi}_{5}} {{\Phi}_{12}^{-}} {{F}_{4R}} {{F}_{4R}} {\overline{F}_{2R}} {\overline{F}_{2R}}+
{{\Phi}_{5}} {{F}_{4R}} {\overline{F}_{2R}} {\overline{\zeta}_{3}} {{Z}_{2}}{\overline{Z}_{4}}+
 {{\Phi}_{5}} {{\zeta}_{3}} {\overline{Z}_{2}} {{Z}_{2}} {{Z}_{4}} {\overline{Z}_{4}}
\nonumber\\ & &\mbox{}+
{{\Phi}_{5}} {{\xi}_{1}} {\overline{Z}_{2}} {{Z}_{2}} {{Z}_{5}} {\overline{Z}_{4}}+
{\overline{F}_{5R}} {\overline{F}_{5R}} {\overline{F}_{3R}}
 {\overline{F}_{3R}} {{\xi}_{1}} {\overline{\xi}_{2}}+
{{F}_{4R}} {{F}_{4R}} {\overline{F}_{3R}} {\overline{F}_{3R}} {{\xi}_{4}} {\overline{\xi}_{2}}+
{{F}_{4R}} {{F}_{4R}} {\overline{F}_{1R}}
{\overline{F}_{1R}} {{\zeta}_{1}} {\overline{\zeta}_{3}}
\nonumber\\ & &\mbox{}+
{{F}_{4R}} {{F}_{4R}} {\overline{F}_{1R}} {\overline{F}_{1R}} {\overline{\zeta}_{1}} {{\zeta}_{3}}+
{{F}_{4R}} {\overline{F}_{3R}} {{\zeta}_{1}} {\overline{\zeta}_{1}} {\overline{Z}_{3}} {{Z}_{4}}+
{{F}_{4R}} {\overline{F}_{3R}} {{\zeta}_{2}} {\overline{\zeta}_{2}} {\overline{Z}_{3}} {{Z}_{4}}+
{{F}_{4R}} {\overline{F}_{3R}} {{\zeta}_{3}} {\overline{\zeta}_{3}} {\overline{Z}_{3}}
{{Z}_{4}}
\nonumber\\ & &\mbox{}+\
 {{F}_{4R}} {\overline{F}_{3R}} {\overline{\zeta}_{3}} {{\xi}_{1}} {\overline{Z}_{3}} {{Z}_{5}}+
{{F}_{4R}} {\overline{F}_{3R}} {{\zeta}_{4}} {\overline{\zeta}_{4}} {\overline{Z}_{3}} {{Z}_{4}}+
{{F}_{4R}} {\overline{F}_{3R}} {{\xi}_{2}} {\overline{\xi}_{2}} {\overline{Z}_{3}} {{Z}_{4}}+
 {{F}_{4R}} {\overline{F}_{3R}} {{\xi}_{3}} {\overline{\xi}_{3}} {\overline{Z}_{3}}
{{Z}_{4}} \nonumber\\ & &\mbox{}+\ {{F}_{4R}} {\overline{F}_{3R}}
{{\xi}_{1}} {\overline{\xi}_{1}} {\overline{Z}_{3}} {{Z}_{4}}+
{{F}_{4R}} {\overline{F}_{3R}} {{\xi}_{4}} {\overline{\xi}_{4}}
{\overline{Z}_{3}} {{Z}_{4}}+ {{F}_{4R}} {\overline{F}_{1R}}
{{\zeta}_{1}} {\overline{\zeta}_{3}} {\overline{Z}_{1}} {{Z}_{4}}+
{{F}_{4R}} {\overline{F}_{1R}} {\overline{\zeta}_{1}}
{{\zeta}_{3}} {\overline{Z}_{1}} {{Z}_{4}} \nonumber\\ &
&\mbox{}+\ {{F}_{4R}} {\overline{F}_{1R}} {\overline{\zeta}_{1}}
{{\xi}_{1}} {\overline{Z}_{1}} {{Z}_{5}}+ {\overline{F}_{3R}}
{\overline{F}_{3R}} {\overline{F}_{2R}} {\overline{F}_{2R}}
{{\xi}_{1}} {\overline{ \xi}_{2}}+ {\overline{F}_{3R}}
{\overline{F}_{3R}} {\overline{F}_{1R}} {\overline{F}_{1R}}
{{\xi}_{1}} {\overline{ \xi}_{3}}+ {{\zeta}_{1}}
{\overline{\zeta}_{3}} {\overline{Z}_{1}} {\overline{Z}_{1}}
{{Z}_{4}} {{Z}_{4}} \nonumber\\ & &\mbox{}+\
{\overline{\zeta}_{1}} {{\zeta}_{3}} {\overline{Z}_{1}}
{\overline{Z}_{1}} {{Z}_{4}} {{Z}_{4}}+ {\overline{\zeta}_{1}}
{{\xi}_{1}} {\overline{Z}_{1}} {\overline{Z}_{1}} {{Z}_{4}}
{{Z}_{5}}+ {{\zeta}_{2}} {\overline{\zeta}_{4}} {{Z}_{2}}
{{Z}_{2}} {\overline{Z}_{4}} {\overline{Z}_{4}}+
{\overline{\zeta}_{2}} {{\zeta}_{4}} {{Z}_{2}} {{Z}_{2}}
{\overline{Z}_{4}} {\overline{Z}_{4}} \nonumber\\ & &\mbox{}+\
{{\zeta}_{4}} {{\xi}_{1}} {{Z}_{2}} {{Z}_{2}} {\overline{Z}_{4}}
{\overline{Z}_{5}}+ {{\xi}_{2}} {\overline{\xi}_{4}}
{\overline{Z}_{3}} {\overline{Z}_{3}} {{Z}_{4}} {{Z}_{4}}+
{\overline{\xi}_{2}} {\overline{\xi}_{4}} {{Z}_{3}} {{Z}_{3}}
{\overline{Z}_{4}} {\overline{Z}_{4}}
\end{eqnarray}

\section{Appendix C: F-- and D--flatness equations}
\newcommand{\FbR}[1]{\overline{F}_{#1R}}
\newcommand{\ksib}[1]{\overline{\xi}_{#1}}
\newcommand{\zb}[1]{\overline{\zeta}_{#1}}
\newcommand{\FR}[1]{{F}_{#1R}}
\newcommand{\z}[1]{\zeta_{#1}}
\newcommand{\ksi}[1]{\xi_{#1}}
\newcommand{\abs}[1]{{\left|{#1}\right|}^2}

The identification of the flat directions in the scalar potential
requires the vanishing of the $F-$ and $D-$terms.  In Section 6,
the complete $F-$flatness conditions  with tree--level
superpotential contributions were presented. Hidden field
contributions are also easily calculated from the superpotential
(\ref{sup}).  Fourth order contributions from both hidden and
observable sectors can also be found from the superpotential terms
(\ref{4nr}) presented in the same section. Higher order terms have
also been calculated and are given in Appendix B. For convenience,
the contributing fifth and sixth order NR superpotential terms are
written  separately in the $w_5''$ and $w_6$ pieces of the NR
superpotential.

{\underline{\it D--flatness}}\\
 The $D$--flatness equations for the non-Anomalous  $U(1)_i$ factors
 are given by
\ba
(D_i): \sum_{\phi_j} Q_j^i |\phi_j|^2&=&0\ ,\;i=1,2,3
\ea
On the other hand,  the Green--Schwarz anomaly cancellation
mechanism in string theory generates a constant Fayet-Iliopoulos
contribution to the $D$--term of the anomalous $U(1)_A$.  This is
proportional to the trace of the anomalous charge over all fields.
To preserve supersymmetry the following $D$--flatness condition
should be satisfied,
\ba
 (D_A): \sum_{\phi_j} Q_j^A |\phi_j|^2&=&-\xi
\ea
where the sum extends over all singlet fields (including the
$SU(4)\times O(4)$ breaking ones) and
$\xi=\frac{3}{8\pi^2}e^{\Phi_D}$. If hidden fields acquire vevs,
they should also be included in the above expressions.

 Taking the combinations $(D_1)$, $\frac{1}{2}\left((D_1)+(D_2)\right)$,
$\frac{1}{2}\left((D_3)-(D_4)\right)$, $\frac{1}{2}(D_4)$ we
obtain
\ba
 \oh\abs{\FbR3}+\abs{\ksi2}-\abs{\ksib2}+\abs{\ksi3}-\abs{\ksib3}
 +{\cal H}_1&=&0\\
 \frac{1}{4}\left(\abs{\FbR2}-\abs{\FbR1}+\abs{\FbR3}+\abs{\FR4}\right)+
 \ \ \
 \ \ \ \ \ \ \ \ \ \ \ \ \ \ \ \ \ \ \ \ \ &&\\
 \mbox{}+\abs{\ksi2}-\abs{\ksib2}+\abs{\bar\Phi_{12}^{-}}-
 \abs{\Phi_{12}^{-}}+\oh\sum_{i=1}^4\left(\abs{\zeta_i}-
 \abs{\zb{i}}\right)+{\cal H}_2&=&0\\
\oh\abs{\FbR3}+\abs{\ksib1}-\abs{\ksi1}+\abs{\ksib4}-\abs{\ksi4}
+{\cal H}_3&=& -\frac{\xi}{3}\\
 \abs{\ksi1}-\abs{\ksib1}+\abs{\bar\Phi_{12}}-
 \abs{\Phi_{12}}+{\cal H}_4&=&\frac{\xi}{2}
\ea
${\cal H}_{1,2,3,4}$ stand for hidden vev contributions. These are
\ba
{\cal H}_1 &=& \frac 12\left(|\bar Z_3|^2 - |Z_3|^2\right)
\nonumber\\ {\cal H}_2 &=& \frac 14\left(|\bar Z_2|^2 +|Z_2|^2
+2\,|\bar Z_3|^2 +|\bar Z_5|^2 -|\bar Z_4|^2 \right)
\nonumber\\
 {\cal H}_3 &=& \frac 14\left(-|\bar Z_2|^2 -|Z_2|^2
+|\bar Z_3|^2 +|Z_3|^2+|\bar Z_4|^2 -|\bar Z_5|^2 \right)
\nonumber\\
 {\cal H}_4 &=& \frac 14\left(|\bar Z_1|^2 +|Z_1|^2+
 |\bar Z_2|^2 +|Z_2|^2-|\bar Z_4|^2 -|Z_4|^2+|\bar Z_5|^2 +|Z_5|^2
\right) \nonumber
\ea
We finally have the  $D$--flatness conditions for the non-Abelian
part of the gauge symmetry. For the $SU(4)\times O(4)$,
\be
\abs{\FbR1}+\abs{\FbR2}+\abs{\FbR3}-\abs{\FR4}=0 \ee
 while in the presence of the hidden non-zero vevs, the
 $SU(8)$ and $U(1)'$ $D-$flatness conditions should also be satisfied
 \ba
 (D_{U(8)}): &\sum_{i=1}^5 \left (|\bar Z_i|^2-|Z_i|^2\right) &= 0,
 \\
(D_{U(1)'}): & \sum_{\chi_j} Q_j' |\chi_j|^2&=0\ .
\ea
where $\chi_j$ stand for all fields carrying $U(1)'$ charge.

 {\underline{\it $F$--flatness}}\\
 {}For completeness, we also write here the $F$-flatness conditions
 including superpotential contributions up to fourth order and hidden
 fields.
Fifth  and sixth order contributions are easily calculated from
the NR--terms presented in Appendix B.
\ba \Phi_2: &  \zeta_i
\bar \zeta_{i}  +
                       \xi_i    \bar\xi_{i} = 0
\label{A 1}\\
\Phi_4: &  \zeta_1  \bar\zeta_{3}  +
                   \bar\zeta_{1} \zeta_3
                     = 0
\label{A 2}\\
\Phi_5: & \zeta_2 \bar\zeta_{4} +
                   \bar\zeta_{2} \zeta_4 = 0
\label{A 3}\\
\Phi_{12}: &
                                 \xi_1 \bar \xi_{4}=0
\label{A 4}\\
\bar\Phi_{12}: &
                     \xi_4   \bar \xi_{1} +
                       Z_3     \bar Z_{3}=0
\label{A 5}\\
\Phi_{12}^-: &  \zeta_i \zeta_i  +
                   \xi_2   \bar \xi_{3}=0
\label{A 6}\\
\bar\Phi_{12}^-: & \bar\zeta_{i}\bar\zeta_{i}  +
                         \xi_3\bar \xi_{2}=0
\label{A 7}\\
\xi_1 : & \phi_2\bar\xi_1
                     + \Phi_{12}\bar\xi_4
            +Z_5\bar Z_5 +
              \bar\zeta_1Z_1\bar Z_1 + \zeta_4Z_2\bar Z_2 = 0
\label{A 8}\\
\bar\xi_1 : &  \phi_2 \xi_1
                      +\bar\Phi_{12}\xi_4=0
\label{A 9}\\
\xi_2:& \phi_2\bar\xi_2 + \Phi_{12}^-\bar\xi_3=0
\label{A 10}\\
\bar\xi_2:& \phi_2\xi_2 +\bar\Phi_{12}^-\xi_3 = 0
\label{A 11}\\
\xi_3:& \phi_2\bar\xi_3+\bar\Phi_{12}^-\bar\xi_2 = 0
\label{A 12}\\
\bar\xi_3:&\phi_2\xi_3+
                      \Phi_{12}^-\xi_2 = 0
\label{A 13}\\
\xi_4:& \phi_2\bar\xi_
         +\bar\Phi_{12}\bar\xi_1 = 0
\label{A 14}\\
\bar\xi_4:& \phi_2 \xi_4
                 + \Phi_{12} \xi_1 = 0
\label{A 15}\\
\zeta_1:& \phi_2\bar\ze_1+\Phi_4\bar\ze_3 + 2\Phi_{12}^-\ze_1 = 0
\label{A 16}\\
\bar\ze_1:& \phi_2\ze_1 +\Phi_4\ze_3 + 2\bar\Phi_{12}^-\bar\ze_1
            +\xi_1Z_1\bar Z_1 = 0
\label{A 17} \\
\ze_2: & \phi_2\bar\ze_2 +\Phi_5\bar\ze_4 + 2\Phi_{12}^-\ze_2
+\frac 1{\sqrt{2}}\bar{F}_{5L}F_{4L}= 0
\label{A 18} \\
\bar\ze_2: & \phi_2\ze_2 +\Phi_5\ze_4 + 2\bar\Phi_{12}^-\bar\ze_2
             + Z_5\bar Z_4/\sqrt{2} = 0
\label{A 19}\\
\ze_3:& \phi_2\bar\ze_3 +\Phi_4 \bar\ze_1 + 2\Phi_{12}^-\ze_3
       + Z_4\bar Z_5/\sqrt{2} = 0
\label{A 20}\\
\bar\ze_3 :&  \phi_2\ze_3 +\Phi_4\ze_1 + 2\bar\Phi_{12}^-\bar\ze_3
           + \bar F_{5R}F_{4R}/\sqrt{2} = 0
\label{A 21}\\
\ze_4: & \phi_2\bar\ze_4 + \Phi_5\bar\ze_2 + 2\Phi_{12}^-\ze_4
         + \xi_1 Z_2\bar Z_2 = 0
\label{A 22} \\
\bar\ze_4: & \phi_2\ze_4 +\Phi_5\ze_2 + 2 \bar\Phi_{12}^-\bar\ze_4 = 0
\label{A 23}\\
Z_1: & \bar\ze_1\xi_1\bar Z_1 = 0
\label{A 24}\\
\bar Z_1:& \bar\ze_1\xi_1 Z_1 = 0
\label{A 25}\\
Z_2 : & \ze_4\xi_1\bar Z_2    = 0
\label{A 26}\\
\bar Z_2 : & \ze_4\xi_1 Z_2    = 0
\label{A 27}\\
Z_3: & \bar\Phi_{12}\bar Z_3 +\bar{F}_{5L}F_{3L}\bar{Z}_4 = 0
\label{A 28} \\
\bar{Z}_3 : & \bar\Phi_{12} Z_3 + F_{4R}\bar{F}_{3R}Z_4 = 0
\label{A 29}\\
Z_4: & \ze_3\bar Z_5/\sqrt{2} + F_{4R}\bar{F}_{3R}\bar Z_3 = 0
\label{A 30}\\
Z_5:& \bar \ze_2\bar Z_4/\sqrt{2} +\xi_1\bar Z_5 = 0
\label{A 31}\\
\bar{Z}_4:& \bar\ze_2 Z_5/\sqrt{2}  = 0
\label{A 32}\\
\bar Z_5: & \ze_3 Z_4/\sqrt{2} + \xi_1 Z_5 = 0
\label{A 33}\\
\FbR5:& \FR4 \zb3=0
\label{A 34}
\end{eqnarray}
\clearpage
 \section{Appendix D: Tree--level  flat directions.}
 We present here the tree--level flat directions of the model.
 As has been explained in Section 6, the solutions are classified
 in four distinct cases according to whether the   singlet vevs
 $\xi_{1,4},\bar{\xi}_{1,4}$ are zero or non--zero. It was shown there
 that only the cases $\xi_1=\xi_4=0$ ( assigned as case ($iii$) in Section 6) and $\bar\xi_1=\bar\xi_4=0$
 (with $\xi_1,=\xi_4 \not= 0$,  referred  as case ($iv$) in the same section) have solutions
 consistent with $F-$ and $D-$ flatness constraints.

 Our analysis proceeded as follows: First we solved  the
 constraints taking into account contributions only from
 the tree--level Yukawa superpotential. An exhaustive analysis
 shows that at tree--level there are 17 solutions for case (iii)
 and 9 solutions for case (iv). These solutions are
 presented in  Tables 9. The  five columns  in the middle
show the fields with zero vevs and the last column the number of
free parameters. For further details in the notation, see
explanation in section 6.  Higher order NR--contributions up to
sixth order, reject several of these cases, resulting to those
 presented in Section 6.

 The complete list of the tree--level solutions  given in Table 9
 is related to flatness constraints involving  fields only from
 the observable sector.  These are easily extended to solutions
 involving hidden fields by using the flatness conditions of
 Appendix C.  Solutions involving hidden field contributions of
 higher NR superpotential terms are more involved and need a
separate treatment.

 \newcommand{\nnu}[0]{\addtocounter{seci}{1}\arabic{seci}}
 \begin{table}[!b]
 \begin{center}
 \setcounter{seci}{0}
 \begin{tabular}{|l|c|c|c|c|c|c|}
 \hline
&$\Phi_{12}s$&$\Phi_i$&$\xi_i,\bar\xi_i$&$\zeta_i,
\bar\zeta_i$&$\bar{F}_i$&f.p.
\\\hline
 $\nnu$&$12,{12}^{-},\overline{12}^{-}$&$2,4,5$&$4,{\bar1},{\bar4}$&$3,
 {\bar3}$&${\bar5}$&$9$\\\hline
 $\nnu$&$12,{12}^{-},\overline{12}^{-}$&$2,4,5$&$4,{\bar1},{\bar4}$&${\bar1},
 {\bar3}$&${\bar5}$&$9$\\\hline
 $\nnu$&$12,\overline{12},{12}^{-},\overline{12}^{-}$&$2,4,5$&$4,{\bar4}$&$3,
 {\bar3}$&${\bar5}$&$9$\\\hline
 $\nnu$&$12,\overline{12},{12}^{-},\overline{12}^{-}$&$2,4,5$&$4,
 {\bar4}$&${\bar1},{\bar3}$&${\bar5}$&$9$\\\hline
 $\nnu$&$12,{12}^{-},\overline{12}^{-}$&$2,5$&$4,{\bar1},{\bar4}$&$3,
 {\bar1},{\bar3}$&$$&$9$\\\hline
 $\nnu$&$12,\overline{12},{12}^{-},\overline{12}^{-}$&$2,5$&$4,{\bar4}$&$3,
 {\bar1},{\bar3}$&$$&$9$\\\hline
 $\nnu$&$12,{12}^{-},\overline{12}^{-}$&$2,5$&$4,{\bar1},{\bar4}$
 &$2,3,{\bar1},{\bar2},{\bar3}$&$$&$8$\\\hline
 $\nnu$&$12,{12}^{-},\overline{12}^{-}$&$2,4$&$4,{\bar1},
 {\bar4}$&$2,3,4,{\bar2},{\bar3},{\bar4}$&${\bar5}$&$7$\\\hline
$\nnu$&$12,\overline{12},{12}^{-},\overline{12}^{-}$&$2,4$&$4,{\bar4}$&
 $2,3,4,{\bar2},{\bar3},{\bar4}$&${\bar5}$&$7$\\\hline
 $\nnu$&$12,{12}^{-},\overline{12}^{-}$&$2$&$2,3,4,{\bar1},{\bar3},
  {\bar4}$&$1,2,3,4,{\bar1},{\bar2},{\bar3},{\bar4}$
 &$\bar 5,\bar 2$&$6$\\\hline
 $\nnu$&$12,\overline{12}^{-}$&$2,4,5$&$2,4,{\bar1},{\bar3},{\bar4}$
 &$1,2,3,4,{\bar3}$&${\bar5}$&$8$\\\hline
 $\nnu$&$12,\overline{12}^{-}$&$2,5$&$2,4,{\bar1},{\bar3},{\bar4}$&
 $1,2,3,4,{\bar1},{\bar3}$&${\bar5}$&$8$\\\hline
 $\nnu$&$12,\overline{12}^{-}$&$2,4$&$2,4,{\bar1},{\bar3},
 {\bar4}$&$1,2,3,4,{\bar2},{\bar3},{\bar4}$&${\bar5}$&$7$\\\hline
 $\nnu$&$12,\overline{12}^{-}$&$2$&$2,3,4,{\bar1},{\bar3},{\bar4}$&
 $1,2,3,4,{\bar1},{\bar2},{\bar3},{\bar4}$&${\bar5}$&$7$\\\hline
 $\nnu$&$12,\overline{12},\overline{12}^{-}$&$2$&$2,3,4,{\bar 1},{\bar3},
 {\bar4}$&$1,2,3,4,{\bar1},{\bar2},{\bar3},{\bar4}$&${\bar5}$&$6$\\\hline
 $\nnu$&$12$&$2$&$2,3,4,{\bar1},{\bar2},{\bar3},{\bar4}$&
 $1,2,3,4,{\bar1},{\bar 2},{\bar3},{\bar 4}$&${\bar5},{\bar 3}$&$7$\\\hline
 $\nnu$&$12,\overline{12},{12}^{-},\overline{12}^{-}$&$2$&$3,4,
 {\bar3},{\bar4}$&$1,2,3,4,{\bar1},{\bar2},
 {\bar3},{\bar4}$&${\bar 5},{\bar 2}$&$8$\\\hline
$\nnu$&$12,\overline{12},{12}^{-},\overline{12}^{- }$&$2,4,5
$&${\bar1},{\bar4}$&$3,{\bar3}$&${\bar5}$&$9$\\\hline
$\nnu$&$12,\overline{12},{12}^{-},\overline{12}^{- }$&$2,4,5$&
${\bar1},{\bar4}$&${\bar1},{\bar3}$&${\bar5}$&$9$\\\hline
$\nnu$&$12,\overline{12},{12}^{-},\overline{12}^{- }$&$2,5$&$
{\bar1},{\bar4}$&$3,{\bar1},{\bar3}$&$$&$9$\\\hline
$\nnu$&$12,\overline{12},{12}^{-},\overline{12}^{- }$&$2,4$&$
{\bar1},{\bar4}$&$2,3,4,{\bar2},{\bar3},{\bar4}$&${\bar5}$&$7$
\\\hline
$\nnu$&$12,\overline{12},\overline{12}^{- }$&$2,4,5$&$2,{\bar1},
{\bar3},{\bar4}$&$1,2,3,4,{\bar3}$&${\bar5}$&$8$\\\hline
$\nnu$&$12,\overline{12},\overline{12}^{- }$&$2,5$&$2,{\bar1},
{\bar3},{\bar4}$&$1,2,3,4,{\bar1},{\bar3}$&${\bar5}$&$8$\\\hline
$\nnu$&$12,\overline{12},\overline{12}^{- }$&$2,4$&$2,{\bar1},
{\bar3},{\bar4}$&$1,2,3,4,{\bar2},{\bar3},{\bar4}$&${\bar5}$&$7$
\\\hline
$\nnu$&$12,\overline{12},\overline{12}^{-
}$&$2$&$2,3,{\bar1},{\bar3},{\bar4}$&$1,2,3,4,{\bar1},{\bar2},
{\bar3},{\bar4}$&${\bar5}$&$7$\\
\hline
 $\nnu$&$$&$$&$2,3,{\bar1},{\bar2},{\bar3},{\bar4}$&$1,2,3,4,
{\bar1},{\bar2},{\bar3},{\bar4}$&${\bar3},{\bar5}$&$7$\\\hline
\end{tabular}
\caption{{ The tree--level solutions to the $F$-- and
$D$--flatness equations. The fields appearing in the table have
zero vevs. In the last column f.p. stands for the number of free
parameters.}}
\end{center}
\end{table}
\end{document}